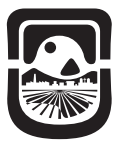

Coniglio

# El Desempleo: estudio de sus causas y posibles soluciones

San Luis. Argentina

octubre de 2012

*Si un hombre intenta diseñar una trampa para ratones mejor, se dirá que es un espíritu emprendedor. Si intenta diseñar una sociedad mejor, se dirá que anda mal de la cabeza.*

*John Kenneth Galbraith*

# RESUMEN

En este trabajo se proponen las siguientes medidas contra el desempleo: a corto plazo, fomentar un mayor ingreso para los sectores más carenciados. Se demostrará que puede pagarse con el aumento de la producción resultante, sin quitarle ingresos a los restantes agentes económicos. A mediano plazo, la creación de empresas ad-hoc para la inversión en proyectos rentables pero de largo aliento. Y a largo plazo, el abandono de los modelos competitivos. Como estas propuestas van contra las actuales corrientes de pensamiento (liberalización, flexibilización, libre mercado, etc.) se demuestran las afirmaciones hechas con minucioso rigor, aún a riesgo de tornar más ardua la lectura del trabajo.

En la parte 1 se explora el problema, y se usa un modelo simple y otros argumentos heurísticos para crear familiaridad con los conceptos macroeconómicos. La parte 2 es un resumen simplificado de la teoría macroeconómica "de libro de texto". Servirá como repaso al lector cuyos conocimientos de economía están oxidados, o como primera aproximación al tema si no los tiene. A la luz de teoría se evalúan políticas económicas para el caso argentino. Se acepta la explicación keynesiana del desempleo (demanda insuficiente), no así el remedio (gasto público). Finalmente en la parte 3 se elaboran y justifican las propuestas.

# INDICE

# 1. VISIÓN PRELIMINAR DEL PROBLEMA

*Agapino Maldonado es medio oficial albañil. Ya no busca trabajo. Vive con un carro de hierro atado a su cuerpo. Ahí guarda todos los despojos que encuentra en la ciudad: vidrios, diarios, chapas, ladrillos. Rastrear en la basura. Arriar el carro. Es el trabajo de los que no tienen trabajo. En la reventa nunca junta más de 150 pesos por mes. En otros tiempos Maldonado iba con su familia al cine Sol de Mayo de Rosario. 'Me gustaban las películas de Kirk Douglas, Burt Lancaster, me gustaban los cowboys'. Algunas veces paseaba con sus hijos en la plaza y hacían un picnic a orillas del río Paraná. 'Siempre pensábamos que nos íbamos a ir de acá. Que íbamos a tener una casa de material. Pero hace tiempo que no podemos pensar en nada. Mi hijo anda descalzo y eso me tiene nervioso´.*[1]

## 1.1 El desempleo ¿consecuencia del progreso?

El desempleo es probablemente el drama social más grande de nuestra época. El hombre con capacidad y voluntad de trabajar siente una frustración enorme cuando se le niega la posibilidad de hacerlo. No sentirse útil, no poder mantener a su familia, lo llevan a incubar resentimientos contra toda la sociedad y sus instituciones, lo que desemboca muchas veces en conductas violentas y delictivas ante la falta de opciones. Pero el desempleado no es la única víctima. Aquellos que conservan sus empleos viven en un estado de temor permanente de ser los próximos despedidos. Los más pudientes intentan aislarse del problema atrincherándose en ciudades fortificadas y con seguridad privada. Con éxito dudoso ya que en algún momento ellos o sus familias tienen que interactuar con el mundo exterior. *"En garitas, rondas y casamatas, más de 8000 custodios protegen los countries del país. La suntuosa y relativa tranquilidad de las villas privadas teme a la marginalidad que la rodea. De allí sin embargo es el personal doméstico para el fin de semana. Simultáneamente, la cantidad de empresas de seguridad creció un 500% en los últimos diez años".* [2]

El problema no se circunscribe a los países periféricos. El Secretario de Trabajo de los Estados Unidos Roberto Reich lanzó recientemente un llamado a sus compatriotas instando a ocuparse del tema con urgencia, ya que no hacerlo implicará costos futuros enormes. Vale la pena citar algunos párrafos:

> *Pensemos por un momento en aquellos que se están quedando atrás. Cuando viajo por Estados Unidos oigo sus temores y veo sus desilusiones. Algunos de ellos están atrapados en islas urbanas de desolación y volencia, cada*

---

1 Reportaje de Marcelo Larraquy, revista NOTICIAS 16/05/96, pag. 112 (fragmento).
2 Revista NOTICIAS, 16/05/96 pag. 106

*día más apartados de la economía general. Otros tienen empleos, pero los empleos no van a ninguna parte. En familias en las que antes bastaba un sueldo, ahora hacen falta dos o tres.*

*Para toda esa gente, el tan elogiado sueño americano parece una mentira cínica. Algunos se entregan al delito. Otros, a demagogos que no tienen dificultades para encontrar blancos a los cuales culpar: los inmigrantes, las madres asistidas por la seguridad social, la acción afirmativa y hasta el propio gobierno.*

*Los exitosos de América no pueden aislarse de una sociedad abandonada. No pagar el precio de preparar a todos nuestros ciudadanos para la nueva economía terminará costando mucho más.*

*California está, como es su tendencia, irrumpiendo primero en el futuro. En 1980 gastaba el 2% de su presupuesto en cárceles y más del 12% en educación superior. Ahora, el presupuesto proyectado del Estado destinará casi el 10% a las cárceles y el 9,5% a educación superior. Para el 2002 el Departamento Correccional se llevará el 18% del presupuesto del Estado; la educación superior tendrá un insignificante 1%. El propio Departamento estima que para el 2027 California tendrá más gente entre rejas que Europa Occidental, Canadá, Nueva Zelanda, Australia y Japón juntos. ¿Tornará eso más productiva nuestra economía? ¿Volverá a nuestra democracia más responsable?*[3]

### 1.1.1 El rol de la tecnología:

Hay factores de producción que con el avance tecnológico se vuelven obsoletos e innecesarios. Un ejemplo claro es el uso del caballo en el cultivo de la tierra: de imprescindible hace 200 años, pasó a tener una importancia secundaria con el uso masivo de maquinaria agrícola.

La sociedad actual asiste a una nueva revolución tecnológica basada en la informática y la robótica, en que las máquinas reemplazan no ya al caballo sino a buena parte del trabajo humano. Esto lleva a los autores más pesimistas a ver el desempleo como una consecuencia inevitable: ahora es el hombre que se vuelve cada vez más prescindible, fenómeno que ya estamos presenciando.

El argumento parece sólido pero es falaz. La gran diferencia es que el hombre **decide** en principio cómo se organiza la sociedad en que vive. Detectando un problema puede encarar soluciones. Por ejemplo y como caso extremo, tiene la **posibilidad** de dejar de fabricar robots o computadoras si estos elementos producen más daños que beneficios. De hecho, hay propuestas bien concretas en esta dirección. *"El Gobierno tiene que impulsar*

---

3. Robert Reich, "Adiós al Sueño Americano", Clarin Económico, 12 de mayo de 1996, pag. 20 (fragmentos).
4. Reportaje a Eduardo Duhalde, "El drama del desempleo", diario CLARIN 26/05/96 pag. 9.

*la obra pública artesanal. Hay que desechar las máquinas por el momento y remplazarlas por el hombre, aún con sistemas de hace 20 ó 30 años. El precio final es el mismo, pero el dinero se reparte entre la gente. No inventamos nada con esto, nos copiamos de lo que hacen otros"*.[4]

Tratemos por un momento de mirar el problema desde una óptica fresca. Imaginemos un marciano que, sin saber nada sobre nuestra sociedad, es colocado en ella como observador. Pronto se dará cuenta del absurdo de la situación: por un lado hay una enorme cantidad de necesidades no satisfechas, por otro maquinaria y gente ociosa. Su propuesta obvia será "pongan a los desempleados a trabajar con las máquinas en las horas libres y repartan lo producido entre todos: habrá más para los pobres y más para los ricos".

Sabemos que la propuesta del marciano es inviable en la práctica: ningún dueño de fábrica abriría sus puertas a desconocidos, y menos para producir lo mismo que él y hacerle competencia. Pero hay una moraleja en el ejemplo. Está claro que el problema no es de origen físico sino institucional. No es el cásico dilema económico de decidir entre usos alternativos de factores de producción escasos. Por el contrario, aquí los recursos (máquinas y hombres) están, la cuestión es encontrar una forma aceptable de incorporarlos a un proceso de producción. Es un problema de **organización social** y no de recursos.

### 1.2 Metodología – Uso de ecuaciones.

En este trabajo usaremos algunas herramientas de matemática elemental, a un nivel accesible a toda persona con estudios secundarios. Mi intención original era prescindir completamente de fórmulas, pero pronto descubrí que eso es imposible, por al menos tres motivos:

Primero: muchos de los problemas tratados son cuantitativos. Un desempleo del 4% es aceptable, uno del 20% no. O bien: las exportaciones favorecen el nivel de empleo, ya que hay que producir los bienes que se exportan. Pero ¿cuánto habría que exportar para eliminar el desempleo? ¿es razonablemente posible alcanzar esta meta? Se necesita un tratamiento que *permita calcular cosas* para responder a estos interrogantes.

Segundo: frecuentemente una acción produce dos efectos opuestos, que pueden cancelarse total o parcialmente. Ejemplo: si se aumenta la alícuota del IVA a 25% ¿sube o baja la recaudación fiscal? Por un lado la recaudación tiende a aumentar por el mayor gravamen a las transacciones realizadas. Pero por otro lado se reduce el número de transacciones: algunas se cance-larán porque ya no son rentables, y otras se harán "en negro" a la nueva tasa. No hay argumento "charlado" que pueda resolver la cuestión, hace falta un tratamiento cuantitativo.[5]

Tercero: trataremos frecuentemente con diez o más variables económicas todas interrelacionadas. El álgebra es una descripción "taquigráfica" para capturar y manejar estas relaciones. Hacer lo mismo en forma verbal llevaría a una maraña de palabras en la que tarde o temprano terminaríamos perdidos.

El discurso del político típico, mientras es opositor, abunda en propuestas sensatas como "incentivar la producción", "fomentar las inversiones", "limitar la especulación", etc. etc. Lo que no dice es *cómo* se logra: qué medidas concretas tomar, y por qué y en qué medida sirven. Cuando ese político llega al gobierno le faltan elementos y termina sin hacer nada de lo prometido. Su discurso fue "guitarreo", no había *sacado las cuentas para ver qué es viable y cómo*. La intención de este trabajo es en lo posible no caer en la misma trampa, queremos dar al lector las herramientas para evaluar cómo, cuánto y por qué. Al respecto es interesante el siguiente párrafo, debido a un Premio Nobel de Economía [6]

> *"Si hoy en día James Clerk Maxwel, uno de los físicos más eminentes del siglo XIX, asistiera a alguna de las reuniones que regularmente celebra la American Physical Society le sería seguramente muy difícil seguir el hilo de las discusiones. Esto sin embargo le habría resultado muy fácil si en lugar de haberse dedicado a la física se hubiera dedicado a la economía, como su contemporáneo John Stuart Mill. La física, a través de la aplicación del método inductivo a los hechos observados y expresados en forma cuantitativa, ha renovado totalmente sus premisas. La economía, en cambio, continúa siendo esencialmente una ciencia deductiva fundada en un grupo estático de premisas, que en su mayoría eran ya conocidas por Stuart Mill y algunas de las cuales encontramos ya en La Riqueza de las Naciones de Adam Smith".*

Compare el lector los avances de una y otra ciencia en el último siglo, y saque sus propias conclusiones. A pesar de lo dicho y para aliviarle el esfuerzo, nosotros usaremos en cada caso la fórmula más sencilla (muchas veces simple proporcionalidad o "regla de tres") que no desvirtúe el fenómeno estudiado, aunque no sea exactamente la fórmula "usual". Pido disculpas a quienes "odian las matemáticas", pero estoy convencido de que para entender los problemas y poder opinar con fundamento hay que "hacer los deberes".

---

5. Este problema específico es tratado por la Curva de Laffer en los textos de economía. Ver por ejemplo el capítulo 6 de The Way the World Works, J. Wanniski, Touchstone Books, Simon & Schuster, New York 1983.
6. W. Leontief, Análisis Económico de Input-Output, Biblioteca de Economía, Ediciones Orbis, Barcelona, 1984, pag. 63

## 1.3 Un ejemplo simple e instructivo.

La única forma de aprender algo complejo es avanzar gradualmente desde casos más simples. Quien aprende a manejar no lo hace en Av. Libertador a las siete de la tarde; empezará en un lugar tranquilo y descampado hasta desarrollar un mínimo de habilidad. El ingeniero dedica años de su formación a estudiar resistencias, condensadores y circuitos simples antes de diseñar una red telefónica urbana. Nosotros vamos a aplicar esta misma metodología, de probada eficiencia, para el problema que nos ocupa. Vamos a estudiar "un sistema económico" aparentemente trivial, el de tres náufragos que llegan a una isla, hasta familiarizarnos con algunos conceptos. Para poder razonar sobre números concretos, hacemos las siguientes hipótesis:

- Dos de ellos, Alberto y Antonio, llegan a nado sin pertenencias. Eduardo por el contrario ha conseguido salvar un bote y elementos de pesca.
- La isla es árida y rocosa, la única actividad económica posible es la pesca.
- Los tres creen en la propiedad privada, la libre empresa, etc. La posesión del bote convierte a Eduardo en empresario, y a sus compañeros de naufragio en asalariados.
- Una hora de pesca rinde en promedio 6 pescados. Cada hombre necesita al menos un pescado por día para sobrevivir, y no puede comer más de 12 porque se llena.
- El pescado no consumido dentro de las 24 hs. se pudre.
- La cantidad de peces en las aguas es enorme, no hay riesgo de que se agoten por la pesca.

Comencemos por señalar que esta sociedad tiene la posibilidad material de satisfacer adecuadamente las necesidades de todos. En efecto, una posible solución al "problema económico de producción y distribución" es la siguiente, que vamos a llamar *Solución solidaria*:

- Alberto y Antonio pescan 3 horas diarias cada uno, lo que da una producción total de 36 pescados por día.
- El salario es de 4 pescados por hora de trabajo. Cada asalariado recibe por lo tanto 12 pescados diarios, quedando otros 12 para el empresario.

Aunque no se tenga una jornada de 8 horas, la situación representa el pleno empleo de esta economía: *al salario vigente nadie busca más trabajo del que ya tiene.*

Hemos supuesto implícitamente una economía de trueque, pero es obvio que nada cambia si introducimos una moneda *siempre y cuando nadie pretenda ahorrar.* Si Eduardo paga a fin del día $4 la hora trabajada y luego vende los pescados a $1 la pieza se obtiene la misma producción y distribución. Una "masa monetaria" de $24 alcanza para que el sistema funcione fluidamente: con esta suma el empresario paga los salarios del día, y recupera el dinero antes de las 24 hs. para el día siguiente. (Volveremos más adelante sobre las complicaciones que introduce el ahorro).

Está claro también que un cambio de precios y salarios en igual proporción no afecta los resultados: daría lo mismo pagar $2 la hora y cobrar $0,50 cada pescado. La única variable relevante es el *salario real* medido en pescados por hora.

Veamos por último un "cambio de escala". Supongamos 100 empresarios con otros tantos botes, y 200 asalariados. El lector puede convencerse fácilmente de que la jornada laboral de 3 horas pagadas a 4 pescados por hora vuelve a dar una situación de pleno empleo, ahora con una producción total 100 veces más grade de 3600 pescados diarios. La producción y consumo *per cápita* no cambian.

**1.3.1 Solución "flexibilizada": [7]**

Supongamos un salario real de 3 pescados por hora en vez de 4. El pleno empleo requiere ahora una jornada laboral de 4 horas para que los asalariados ganen los mismos 12 pescados diarios que les interesa obtener. Haciendo números se ve inmediatamente que con tal jornada el sistema no está en equilibrio[8]. Si Alberto y Antonio trabajan 4 horas cada uno, la producción es 48 pescados. Ellos reciben 24 como salario, Eduardo consume 12, y los otros 12 se pudren. **Hay sobreproducción.**

En consecuencia, al día siguiente se los contrata por menos horas, digamos 3. Producción: 36 pescados. Salarios 18 (6 horas-hombre a 3 pescados la hora), consumidos por Eduardo 12, sobran 6 que se pudren. **Aún hay sobreproducción.**

Al otro día se los contrata por 2 horas. Producción 24 pescados, salarios 12, consumidos por Eduardo otros 12, no sobra ninguno. **Hemos encontrado el nuevo punto de equilibrio**. La producción cayó un 33% respecto al caso solidario, el consumo de los asalariados un 50%, y hay un desempleo del 50% (Alberto y Antonio trabajan la mitad de las horas que quisieran).

---

7. Siguiendo la costumbre en boga, usamos el eufemismo flexibilización para designar una baja del salario real.
8. Se dice que una economía está en equilibrio si los valores de sus variables pueden repetirse durante sucesivos períodos de tiempo, como es el caso de nuestra "solución solidaria".

Por supuesto podríamos volver a hacer las mismas observaciones sobre uso de una moneda y cambios de escala, nada cambiaría mientras el salario real se mantenga en 3 pescados por hora.

### 1.3.2   Solución "sindical":

Vayamos al otro extremo. Supongamos que de alguna forma Alberto y Antonio consiguen un salario de 5 pescados por hora (digamos, formando un sindicato poderoso). Ahora ganan sus 12 pescados diarios con una jornada laboral de 2hs. y 24 minutos (el salario es 1 pescado cada 12 minutos de trabajo), más allá de lo cual no les interesa trabajar. La producción es de 28,8 pescados diarios[9] de los cuales 24 van para Alberto y Antonio, y 4,8 para Eduardo.

La producción cayó 20% respecto del caso solidario, el consumo empresarial 60%. Este caso es el simétrico del "flexibilizado": ahora es el empresario quien quisiera vender más de su factor de producción (el uso del bote) y no encuentra mercado para hacerlo. Sería lógico llamar a este estado de cosas *desempleo del capital.*

### 1.3.3   El salario real como variable

Hemos determinado las diversas variables del modelo para tres salarios diferentes. Siguiendo igual procedimiento podemos estudiar otros valores y construir una tabla o un gráfico donde se muestre la producción y los consumos per cápita en función del salario. Las cuentas están hechas en el apéndice 1.A, presentamos aquí simplemente los resultados en el gráfico que sigue. La línea llena da tanto la producción total (escala de la izquierda) como la jornada laboral o "nivel de empleo" (escala derecha).

---

9. En un día cualquiera obviamente se saca un número entero de pescados, pero este es un detalle técnico sin importancia: aquí hablamos de un promedio. Por ejemplo, podrían pescarse 29 pescados 4 días de cada 5, y 28 el restante.

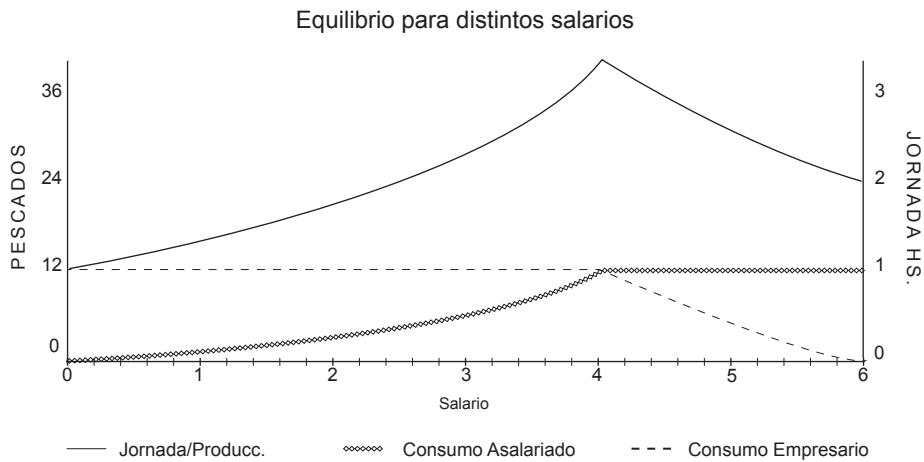

Equilibrio para distintos salarios

Figura 1.1

## 1.4 Primeras enseñanzas del ejemplo

Se ve en el gráfico previo que la producción, y el nivel de empleo que le es proporcional, alcanzan un valor máximo para cierto salario (w=4 con los números que hemos supuesto), o sea para cierta *distribución de la producción entre los diferentes agentes económicos*. Podemos llamar a esta distribución *equilibrada* en el sentido de que da a cada uno según sus deseos de consumo. Cualquier apartamiento de tal distribución, ya sea en detrimento de los asalariados o del empresario, reduce la producción y el empleo. La razón por la cual esto ocurre es clara en el ejemplo: si un día cualquiera *alguien* gana más de lo que puede consumir, hay sobreproducción y eso lleva a bajar la actividad al día siguiente. La producción baja hasta el nivel en que *quien más gana* recibe una cantidad igual a su consumo.[10]

Obviamente nuestro simple ejemplo no describe una economía real, pero hay conclusiones lógicas que si podemos sacar. Se escucha a diario en casi todos los medios masivos la siguiente afirmación:

**"La flexibilización laboral va a aumentar el nivel de empleo."   Afirmación (a)**

Esto se enuncia como una supuesta *Ley General de la Economía* que no necesita demostración (por lo menos nadie se molesta en darla). Los defensores de (a) ignoran alegremente la evidencia empírica de la última década,

---

10. Más adelante veremos cómo se modifica la situación cuando hay ahorro.

en que el salario real cayó varias decenas de puntos mientras el desempleo aumentó enormemente.

Si (a) fuera en efecto una ley general debería aplicarse en todos los casos, incluida nuestra isla para cualquier *w*. Ello no ocurre, por lo tanto *(a) no es una ley general.* Y si no es una ley general, el rigor lógico requiere que su validez sea *demostrada* en cada caso particular. Quienes afirman (a) **tienen la obligación** de explicar por qué habría de valer para la economía argentina algo que no vale en un caso mucho más simple.[11]

Lo que sucede es que (a) está basada en una conocida falacia, la *falacia de composición*. Otro premio Nobel, Paul Samuelson, la define de la siguiente manera [12]

**Falacia de composición: consiste en suponer que lo válido para cada parte es automáticamente válido para el todo.**

Y pasa a ilustrar la idea con el siguiente ejemplo: un espectador ve mejor el desfile si se para en puntas de pie (verdadero). Todos los espectadores ven mejor el desfile en puntas de pie (falso, falacia de composición).

Volvamos a la isla y veamos concretamente dónde está el error. Supongamos como punto de partida la solución flexibilizada con salario *w = 3* y jornada *j = 2* horas. . Un asalariado, digamos Alberto, efectivamente puede conseguir más horas de trabajo si se ofrece a un salario menor, por ejemplo a *w = 2,5*. Al nuevo salario a Eduardo le conviene despedir a Antonio y darle todo el trabajo a Alberto. Hasta aquí todo bien, (a) funciona *para un asalariado particular.* Pero ahora viene el problema: creer que Antonio también va a trabajar más horas a *w = 2,5* es caer en la falacia: al nuevo salario ambos trabajan menos (concretamente, *j = 1h 42'* y fracción si se sacan las cuentas).

En general es cierta a nivel macroeconómico (esto es, para una parte pequeña de la economía) la afirmación siguiente:

**"Al bajar el precio de un bien, la cantidad demandada sube". Afirmación (b)**

---

que aplicada al caso particular de la mano de obra parecería validar (a), pero al tratar de aplicarla a nivel macroeconómico (para la totalidad) se comete la falacia de composición. Para una discusión más detallada, ver el apéndice 1.B.

## 1.5 Ampliando el modelo

El ejemplo de la isla, con todas sus simplificaciones, tiene la rara virtud de permitir la visualización y comprensión intuitiva de algunas actividades económicas difíciles de aprehender en casos más complejos. Vamos a seguir explotándolo un poco más.

**1.5.1 Competencia, economía "de mercado".**

Consideremos la competencia originada en la llegada de un tercer naufragio, Ariel, que tampoco trae "capital". Supongamos inicialmente el caso solidario $w = 4$. El recién llegado necesita empleo. Para lograrlo ofrece sus servicios por $w=3,5$, con lo cual desplaza a uno de los pescadores previamente empleados, digamos Alberto. Este, obviamente, necesita trabajar y baja también su pretensión al nuevo salario, despazando a Antonio que seguía ganando 4. Ahora Antonio debe bajar su salario *por debajo de w=3,5* para ser tomado, y así siguiendo. A cada nueva baja del salario, mayor "beneficio patronal" por hora de trabajo, y por lo tanto menos horas trabajadas por cada uno para producir los 12 pescados para Eduardo. El proceso puede terminar de dos formas: o bien los pescadores se agremian y fijan un salario común y un reparto del trabajo disponible (fin de la competencia), o bien alguno cae debajo del nivel de subsistencia (1 pescado diario) y entonces "patea el tablero" con alguna salida violenta.

Lo interesante es que, incorporado el recién llegado, podrían haber negociado una nueva *solución solidaria* con los siguientes valores: salario real *w=4,5* pescados por hora, jornada laboral *j= (8/3 hora= 2h 40'*, producción de cada pescador: 16 pescados diarios, de los cuales se lleva 12 como paga y deja 4 para Eduardo, cuyo consumo es entonces *E= 3 x 4= 12*. Nuevamente es una situación óptima en el sentido de que cada uno vuelve a tener todo lo que quiere. Nótese que al aumentar la oferta de mano de obra, el óptimo requiere mayor salario y menor jornada laboral. **Exactamente al revés del resultado competitivo.** ¿Cómo? ¿No era que la competencia siempre es beneficiosa para la sociedad? ¿No se afirma permanentemente que

**"La libre competencia maximiza el bienestar general"       afirmación (c)?**

El ejemplo muestra claramente que ésta tampoco es una ley general, aunque pueda ser válida en muchos casos. Y no lo decimos sólo nosotros, oigamos la opinión de un especialista[13]

**Una mala nomenclatura (óptimo según Pareto) en la literatura, junto con mucho descuido en los libros de texto, a menudo desorienta a la gente haciéndole creer que existe algún teorema que afirma que un equilibrio competitivo es socialmente óptimo. No existe tal afirmación.**

Se deja como ejercicio examinar el problema simétrico: el recién llegado trae un bote y compite con Eduardo como empresario. Las conclusiones son las mismas: a) la competencia, si no se para, lleva al colapso del sistema, y b) existe una solución solidaria en que las variables se desplazan en sentido inverso al "esperado".

### 1.5.2 Ahorro, inversión

Veamos ahora la discusión pendiente sobre las consecuencias del ahorro, para lo cual suponemos una economía con dinero. Comenzamos con una observación sencilla pero importantísima. Si la cantidad total de dinero es fija, **por cada individuo que ahorra debe haber otro que "desahorra" igual monto.** Si yo termino el día con $1 más en mi bolsillo que esta mañana, ese $1 está faltando en algún otro bolsillo, alguien "desahorró". En una economía real Ud., yo y el vecino de enfrente, si tenemos suficientes ingresos, podemos ahorrar $100 cada uno este mes; no hay restricciones para el ahorro individual. Concluir de ahí que todos podemos ahorrar $100 mensuales es cometer nuevamente la falacia de composición. Que todos los individuos lleguen a fin de mes con $100 más que al principio es imposible porque la cantidad de dinero circulante es fija, lo que uno tiene de más le tiene que estar faltando a otro.

---

13. Frank Hahn, La Crisis en la Teoría Económica, Ediciones El Cronista Comercial, capítulo VIII "La teoría del equilibrio general", Pag. 179. Se trata de la traducción The Crisis in Economic Theory, Basic Books, NY 1981, obra que reúne las opiniones de 12 autores de diversas tendencias.

Hecha esta aclaración, volvamos a nuestra isla. Tomamos como punto de partida la "solución solidaria" (cualquier otra daría lo mismo para el análisis). Un buen día Alberto decide ahorrar siguiendo la siguiente estrategia: de su paga gastará todos los días hasta $8 en pescados, y guardará el resto. Veamos la contabilidad de nuestros amigos ese día:

| Alberto | | | Antonio | | | Eduardo | | |
|---|---|---|---|---|---|---|---|---|
| Sueldo | | 12 | Sueldo | | 10 | Ventas | | 20 |
| Compras | 8 | | Compras | 10 | | Sueldos | 24 | |
| Ahorro | 4 | | | | | Déficit | | 4 |

Alberto ha ahorrado efectivamente sus $4, pero a expensas de un desahorro (totalmente involuntario) por parte de Eduardo que pagó $24 en sueldos y sólo recuperó $20 por ventas. Obviamente esta situación no le satisface, no quiere poner dinero de su bolsillo para pagar sueldos. Dado que "las ventas están flojas", al día siguiente reduce la jornada laboral en 20 minutos, justo lo necesario para pagar sólo $20 en salarios. Al segundo día tenemos:

| Alberto | | | Antonio | | | Eduardo | | |
|---|---|---|---|---|---|---|---|---|
| Sueldo | | 10 | Sueldo | | 10 | Ventas | | 18 |
| Compras | 8 | | Compras | 10 | | Sueldos | 20 | |
| Ahorro | 2 | | | | | Déficit | | 2 |

Eduardo sigue en déficit, así que reduce la jornada laboral en otros 20 minutos[14]. Con esto se alcanza un nuevo equilibrio, quedando la situación así:

| Alberto | | | Antonio | | | Eduardo | | |
|---|---|---|---|---|---|---|---|---|
| Sueldo | | 8 | Sueldo | | 8 | Ventas | | 16 |
| Compras | 8 | | Compras | 8 | | Sueldos | 16 | |
| Ahorro | 0 | | | | | Déficit | | 0 |

El intento de ahorrar llevó a un nuevo equilibrio **en que el ingreso diario del "ahorrista" bajó lo suficiente como para frustrar su intento**. Y de paso arrastró al otro pescador en la caída salarial.

---

14. Su argumento sería; disminuí la carga salarial en $4 y bajé el déficit en $2, si vuelvo a hacer lo mismo bajo el déficit a cero.

Para quien quiera agilizar su comprensión del problema, le propongo como ejercicio ver qué sucedería con las siguientes variantes:

- Si desde el primer día Eduardo trata de eliminar su déficit bajando el precio del pescado para vender más, en lugar de reducir la jornada laboral.
- Si quien decide ahorrar es Eduardo en ves de Alberto.
- Si hay varios empresarios y asalariados (digamos 3 y 6) y uno de ellos decide empezar a ahorrar (este es más difícil, deberá hacer suposiciones adicionales razonables sobre el comportamiento de los diversos individuos).

¿Significa todo esto que el ahorro es imposible en un estado de equilibrio? No, significa solamente que para que alguien pueda ahorrar, algún otro debe **desahorrar voluntariamente**, es decir tomar dinero prestado, lo que hasta aquí no sucede en nuestro ejemplo. El caso más común de "desahorro voluntario" es la *inversión*: alguien toma dinero prestado para financiar hoy un emprendimiento que le dará ganancias futuras. Veamos cómo funcionaría. Volvamos al día inicial, y supongamos que Antonio quiere armar una empresa de espectáculos amaestrando un delfín. Necesita cuatro pescados diarios para premiarlo por lo que aprende, y busca un capitalista que le preste los fondos necesarios. Está dispuesto a endeudarse en $4 diarios (desahorrar) como inversión en su empresa. Convence a Alberto de que le preste sus ahorros en lugar de guardarlos en el colchón. La nueva contabilidad del día uno queda así:

| Alberto | | | Antonio | | | Eduardo | | |
|---|---|---|---|---|---|---|---|---|
| Sueldo | | 12 | Sueldo | | 12 | Ventas | | 24 |
| Compras | 8 | | Compras | 16 | | Sueldos | 24 | |
| Ahorro | 4 | | Préstamo | 4 | | Déficit | | 0 |

Ahora no hay desequilibrio, lo mismo puede pasar en días sucesivos sin que nadie desee alterar su conducta. Esto se ha conseguido porque los préstamos demandados igualan exactamente el ahorro. En otras palabras, el ahorro neto voluntario (contando como negativo el desahorro) es cero.

Se deja como ejercicio (y es altamente recomendable hacerlo) encontrar el equilibrio si Antonio sólo invierte $2 diarios en su empresa mientras Alberto pretende ahorrar $4 (ayuda: al llegar al equilibrio debe tenerse nuevamente Ahorro=Inversión como en los dos casos tratados).

### 1.5.3 Otras posibles generalizaciones

Abandonamos aquí la isla, aunque aún se le puede sacar mucho jugo como herramienta didáctica. Para el lector curioso, he aquí un par de posibilidades que puede interesarle explorar:

- Agregar una nueva profesión (digamos maestros) y un "gobierno" que les paga cobrando impuestos a la producción de pescados.
- Agregar "comercio exterior". Se descubre en el extremo opuesto de la isla otra colonia de náufragos cuya industria es el cultivo de papas. Efectos sobre producción y empleo en economía de trueque y monetaria. Equilibrio de la paridad cambiaria (suponer cantidades fijas de ambas monedas, y que ninguna de las dos colonias quiere agotar su stock de dinero). Este es bastante difícil y probablemente no pueda resolverse en detalle, pero reflexionar sobre el comercio exterior en un caso "trivial" ayuda a comprender el mundo real.

En los párrafos siguientes cambiaremos de enfoque. En lugar de analizar en todo detalle un modelo simple, como se ha hecho con la isla, pasamos a considerar la economía real. Dada su complejidad, será mucho menos lo que podamos decir con certeza. Veremos sin embargo que se corroboran los resultados hasta aquí obtenidos, proveyendo una especie de "movimiento de pinzas" conceptual para ir delimitando las ideas válidas.

## 1.6 La economía real: cómo provocar desempleo.

Uno de los recursos más fructíferos en ciencias es la *experimentación*, que permite estudiar el efecto de variaciones controladas de alguna variable sobre el comportamiento de un sistema. En Economía tal recurso no está disponible; hay razones prácticas y éticas que impiden experimentar con una sociedad. Lo que sí podemos hacer es analizar experimentos conceptuales (los llamados *Gedankenexperimente* de las ciencias físicas) donde imaginamos las consecuencias de determinadas acciones. Esta sección propone un tal ejercicio, que se revelará muy fructífero. Para aprender cómo salir de una recesión, veamos cómo se llega a ella en primer lugar.

Tomemos el siguiente escenario. La economía argentina ha funcionado con eficiencia escandinava durante algún tiempo. El PBI es de u$s 350.000 millones anuales, con un ingreso *per capita* de u$s 10000. El desempleo es del 3% y el sueldo promedio u$s 1200 mensuales, un poco superior a la canasta familiar. La jubilación tipica es de u$s 900. Hay un nivel de inversión razonable y se exporta por u$s 35000 millones anuales (10% del PBI).

Un buen día, por razones que no vienen al caso, se promulga una ley con el extraño nombre de "Ley Ozairalas" (antisalariazo) según la cual se bajan todos los sueldos a la mitad. No hace falta ser economista para entender en líneas generales las consecuencias de este cambio. Los eventos que se desencadenan serán aproximadamente los siguientes:

- a) En una primera instancia, unos 30.000.000 de personas entre asalariados, jubilados y sus grupos familiares, se encuentran con la mitad del dinero habitual. Por lo tanto suprimen gastos no esenciales: libros, diarios, espectáculos, vestimenta, equipamiento del hogar, esparcimiento, auto, etc.

- b) Simultáneamente las empresas se encuentran con un excedente de dinero al pagar sueldos menores. Una empresa con 100 operarios "ahorrará" unos 100*1.200/2 = 60.000 u$s (antes pagaba 100 sueldos medios o u$s 120.000, ahora sólo la mitad).

- c) En una segunda instancia, en cuestión de semanas y como consecuencia de a) se produce una grave crisis en las industrias afectadas: editoriales, cines, teatros, textiles, electrodomésticos, clubes, automotrices, etc. La caída drástica de las ventas las obliga a despedir gente, o lisa y llanamente a cerrar. El hecho de que los empresarios ganen ahora más no corrige la escasa demanda: por mucho que gane, un empresario no usará 20 butacas del cine ni se pondrá 14 camisas.

- d) En una tercera instancia, los nuevos despedidos del punto c) dejan a su vez de consumir. Las ventas caen ahora para todos los bienes: jabón, chocolate, lentejas, sillas, colchones, lácteos, fiambres, etc. Ya no son sólo los bienes prescindibles los que se venden menos. La crisis se propaga, todas las industrias tienen que reducir personal ante la caída de las ventas.

- e) Los despidos en las restantes industrias llevan a una nueva caída del consumo, y así siguiendo. Por el momento no tenemos herramientas analíticas para calcular hasta dónde sigue este proceso (las adquiriremos en la parte 2), pero está clarísimo que la *"Ley Ozairalas" produce drásticos aumentos del desempleo y caídas en la producción.*

- f) Puede suceder que ante la caída de ventas los empresarios reduzcan sus precios. Un poco de reflexión muestra que ello morigera pero no elimina el problema. En primer lugar y dada la mentalidad empresaria, es poco probable que el menor costo se traslade íntegramente al precio. Y segundo, aún así no se restablece el poder adquisitivo original ya que los sueldos representaban sólo parte del costo. Un ejemplo aclarará este último punto. Supongamos que inicialmente el costo de un Kg. de harina era de $1, con $0,50 en salarios y $0,50 en beneficios del agricultor, del dueño del molino, etc. Después de Ozairalas el costo del Kg. es

$0,75: $0,25 en salarios y $0,50 en beneficios. El precio cae un 25% y los sueldos 50%, igual hay pérdida de poder adquisitivo, aunque menor.

- g) La situación expuesta generará todos los perniciosos efectos sociales conocidos: aumento de la marginalidad, miseria, delincuencia, auge de la "timba" como esperanza de salvación individual entre los asalariados, fin de la solidaridad al competir por los pocos empleos disponibles, búsqueda de negocios con el Estado (que sigue siendo un consumidor solvente y muchas veces generoso) por parte de los empresarios, por medios lícitos o no tanto (coimas), etc. etc.

### 1.6.1 Conclusiones preliminares

Si se ven analogías entre el escenario antes descripto y la realidad cotidiana, no es coincidencia. Según la teoría (¿ingenua?) que los medios nos presentan a diario, a menor costo de la mano de obra debería aumentar el empleo. Eso puede ser cierto para una industria aislada, pero nuestra discusión muestra claramente que a nivel global ocurre todo lo contrario. Nótese que este ejemplo, basado en una economía con todas las complejidades del caso real, corrobora totalmente lo encontrado para la isla: la caída del salario real o "flexibilización" reduce el nivel de empleo y la producción total. Cuanta más flexibilización, más se refuerza el efecto.

El hecho de encontrar respuestas concordantes al enfocar el problema desde extremos opuestos es buena señal, sugiere que vamos por buen camino. Y entendiendo cómo se produce una recesión sabemos también cómo salir de ella: hay que revertir la baja de salarios. En el caso de la isla podía hacerse sin perjudicar en absoluto al empresario (recordar la "solución solidaria"). Veremos más adelante (parte 3) si lo mismo pasa en el caso real.

### 1.6.2 Relevancia del ejemplo

Hay un detalle, y es que nunca se promulgó una Ley Ozairalas. Por lo tanto ¿Qué tiene que ver todo lo discutido con el caso real? Creo en base a mi experiencia y recuerdos personales, que si bien nunca existió tal ley, el efecto Ozairalas se fue produciendo poco a poco como consecuencia del deterioro del salario real durante largos períodos inflacionarios. Y ley o no ley, el estado final alcanzado es el mismo. Si bien carezco de números confiables para demostrarlo, permítaseme citar algunas cifras que personalmente me constan.

- A fines de la década del 40 mi padre era un modesto oficinista con 300$ "moneda nacional" de sueldo. Probablemente ese era el salario mínimo de la época. El boleto de colectivo costaba 10 centavos, el helado "sándwich" (bloque de helado entre dos obleas) otro tanto, y el "café con leche

completo", con medias lunas y dulce, 40 centavos. Con un sueldo se pagaban 3000 boletos o helados, o 750 café con leche completos. Digamos que hoy esos productos valen $0,50, $1 y $3 respectivamente. De haberse mantenido el poder adquisitivo, el sueldo actual debería ser $1500, $3000 y $2250. Dudo que un trabajo similar se pague hoy a más de $600.

- En los años 60 me pagué mis estudios con dos ayudantías "de segunda" en la Universidad, lo que ahora se llama "ayudante alumno". Pude alquilar casa, comprar muebles, casarme, mantener y costear los estudios de mi mujer, salir de vacaciones dos veces por año, ir al cine o al Colón tres o cuatro veces por semana con cena afuera incluida, y encima ahorrar, sin otra asistencia económica. Hoy un ayudante alumno gana $70 mensuales (no es un error, son $70!!).

- A comienzos de los '70 vivía cómodo en los EE.UU., con mujer e hijo, con una beca de u$s 800. Pagaba u$s 89 de alquiler por un departamento modesto, y luego u$s 205 por uno muy confortable en un suburbio. Una casa de suburbio costaba u$s 40.000, un almuerzo en el "shopping" u$s 2,50, un paquete de cigarrillos u$s 0,29 (comprando por cartón en el supermercado). Hoy los precios de los mismos ítems se han multiplicado por 4, 4, 10, 4 y 5 respectivamente. La misma beca son unos u$s 1500, no llega al doble. *El efecto Ozairalas no es patrimonio argentino!!*

El lector podrá ampliar la lista de ejemplos con otros de su propia cosecha, o los más jóvenes consultar a padres y/o abuelos, o tal vez recordar algunos precios de comienzos de la "convertibilidad" y comparar con los actuales. Encontrará sin dudas evidencias de un fuerte "efecto Ozairalas" en el transcurso de los años, de modo que nuestra discusión **es relevante** para entender lo que está pasando en Argentina y mundialmente.

### 1.6.3 Un gráfico ilustrativo

Podemos poner los resultados previos en un gráfico, que nos ayuda a resumir y fijar las ideas. Representemos el nivel de empleo, en horas hombre por mes o cualquier otra unidad conveniente, sobre el eje horizontal. Representemos la producción total del país en el mismo período, medida en dinero a precios de mercado, sobre el eje vertical. En el corto plazo (unos pocos meses) podemos olvidarnos de la acumulación de capital y los cambios tecnológicos; la producción será mayor cuanto mayor sea el empleo. Sin entrar en sutilezas, supongamos simple proporcionalidad o "regla de tres": si 1.000 horas-hombre producen $10.000, 1.200 horas hombre producen $12.000, etc. Obtenemos el gráfico de la figura 1.2

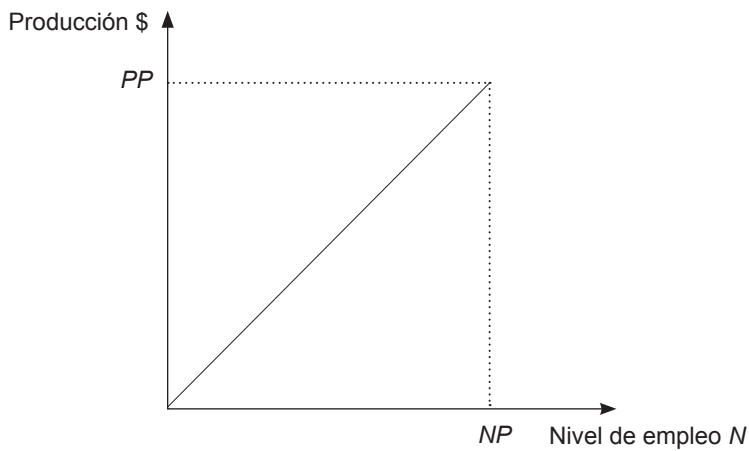

Figura 1.2

El punto *NP* corresponde al pleno empleo (100% de la población activa tiene trabajo), por lo que la gráfica "se corta" allí, no se puede tener empleo ni producción mayores porque no hay bastante gente. La producción *PP* correspondiente se llama *potencial productivo* de la sociedad, es lo más que puede producirse con los recursos disponibles.

Veamos ahora cuánto pueden vender las empresas a cada nivel de empleo para un salario dado. Podemos afirmar: primero, que aún si nadie tuviera trabajo habría ventas, ya que hay gente con ahorros. La gráfica de ventas por lo tanto corta al eje vertical por encima del cero. Segundo, las ventas crecen con el nivel de empleo: quien trabaja gana plata para gastarla. Pero crecen menos rápidamente que la producción, ya que cada asalariado cobra menos $ de los que produce (obvio, parte del valor producido es beneficio del empresario, de lo contrario nadie pondría una empresa). Y parte de lo que gana, la gente lo ahorra. Nuevamente sin entrar en sutilezas, trazamos para las ventas una línea recta como se ve en la figura 1.3

El nivel de empleo de equilibrio es aquel que igual producción y ventas, es decir el punto Neq del gráfico. En efecto, si un mes hay empleo mayor a *Neq,* la producción supera a las ventas acumulándose el excedente como exceso de stocks. Los empresarios no quieren acumular stocks que no venden y bajan la producción (y por lo tanto el empleo) al mes siguiente. Análogamente, si *N* es inferior a *Neq* se vende más de lo que se produjo despoblando los stocks de reserva. Al mes siguiente los empresarios aumentan la producción para repoblarlos.

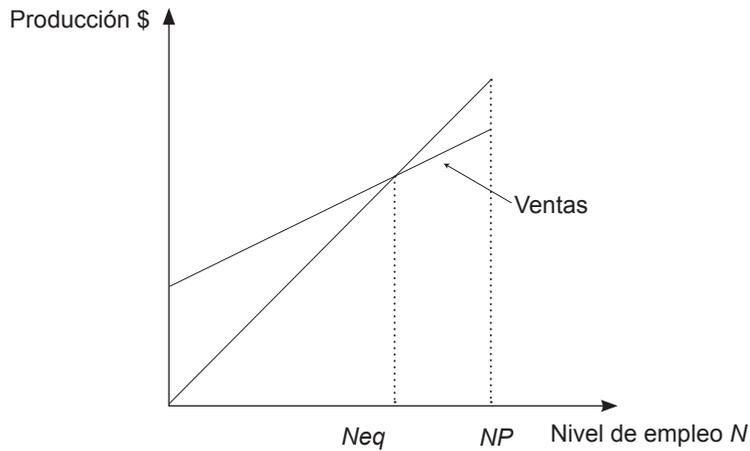

Figura 1.3

No hay ninguna razón *a priori* para que el empleo de equilibrio *Neq* coincida con el pleno empleo *NP*, más aún, sería una increíble coincidencia. Supongamos *Neq* < *NP* al salario vigente (recordar que el gráfico se hizo para un nivel de salarios fijo). ¿Qué sucede si se flexibiliza el salario, es decir si dejamos que empleados y desocupados compitan por el trabajo existente? Se produce una baja de los salarios, con lo cual la recta de ventas se hace más horizontal tal como se ve en la figura 1.4. Se ve claramente que al menor salario "flexibilizado" hay menor empleo, y que lo necesario para desplazar *Neq* hacia *NP* sería la medida opuesta, recomponer el salario real. Algo que la competencia entre asalariados acuciados por el fantasma del desempleo no puede lograr nunca.

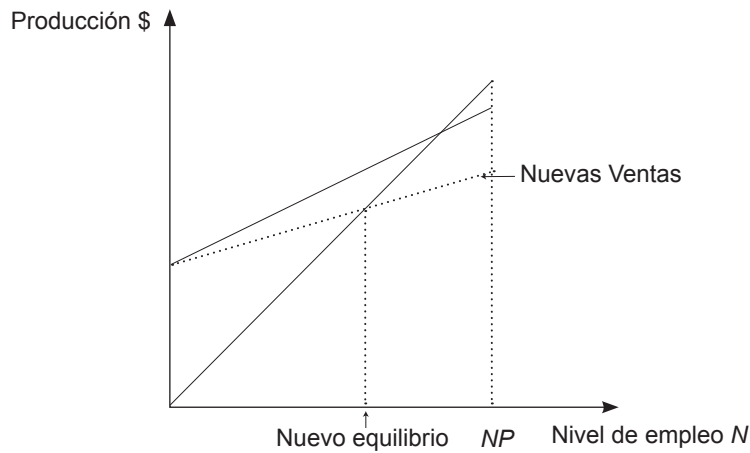

Figura 1.4

El argumento presentado da la idea básica pero aún carece de rigor ¿qué significan exactamente "producción" y ventas" en el gráfico? ¿Cómo se calculan a partir de datos reales? ¿Qué pasa si hay exportaciones? etc., etc. La parte 2 de este trabajo responde a tales interrogantes. Veremos en particular la versión usual del mismo gráfico, en que nuestra producción se sustituye por el producto bruto interno PBI y las ventas por una función de consumo, ambas cantidades rigurosamente definidas.

## Apéndice 1.A – Solución analítica de la economía de isla

Para encontrar los puntos de equilibrio de la "economía de isla" para distintos valores del salario real $w$ podríamos proceder por tanteos como en la sección 1.3, pero es más expeditivo usar un procedimiento analítico. Sea $w$ el salario real (en pescados por hora), y $j$ la duración de la jornada laboral. Tenemos entonces una producción total *P = 12 \* j* (un pescador saca *6 \* j* pescados por día, y hay dos pescadores). De la producción total $A = j * w$ pescados van a cada asalariado, quedando E = 12 \* j - 2 \* j \* w = 2 \* j \* (6-w) para el empresario.

La solución solidaria se caracteriza por dar igual número de pescados a cada uno. Poniendo *A = E* y usando las expresiones previas de *A y E* obtenemos la ecuación *j \* w= 2 \* j \* (6-w)*, en el cual *j* se cancela para dar *w= 2 \* (6-w)*, de donde se despeja *w= 4*. Usando el hecho de que *A= 12= j \* w* se encuentra *j = 3*, y sustituyendo estos *j* y *w* en las expresiones de *P* y *E* encontramos:

**Solución solidaria:**

  *J = 4*    *P = 36*    *E = 12*    *A = 12*

coincidente con lo dicho en la sección 1.3. En la solución flexibilizada, válida para salarios inferiores al solidario, el empresario recibe 12 pescados diarios. Poniendo *E= 2 \* j \* (6-w)= 12* despejamos *j= 6 / (6-w)* fórmula que da *j* para cada valor de w. Reemplazando este j en las expresiones de *P* y *A* tenemos las siguientes expresiones:

**Solución para salario flexibilizado w<= 4:**

  *j = 6 / (6-w)*    *P = 12 \* 6 / (6-w)*    *E = 12*    *A = 6 \* w / (6-w)*

En la solución "sindical" *w >= 4* son los asalariados quienes reciben 12 pescados: *A = j \* w = 12* de donde despejamos *j = 12/w*, valor que sustituido en las expresiones de *P* y *E* da

**Solución sindical w>= 4:**

*j = 12 / w*     *P = 12 \* 12 / w*     *E = 2 \* 12 \* (6-w) / w*     *A = 12*

Nótese que para *w = 4* puede usarse cualquiera de los tres grupos de fórmulas con resultados idénticos. Con estas fórmulas y cualquier planilla de cálculo se construye el gráfico de la figura 1.1

Ejercicio: resolver el mismo problema para los siguientes casos: a) un empresario y tres asalariados, b) dos empresarios (cada uno con su bote) y dos asalariados, c) dos empresarios y tres asalariados.

Si el lector es incapaz de resolver estos ejercicios se le recomienda repasar su matemática del secundario, o sacará poco provecho del resto del material presentado.

Apéndice 1.B – La oferta y la demanda.

Consideremos un bien en particular, por ejemplo el pan. Resulta razonable suponer que:

- Cuanto mayor es su precio, menor será la demanda, es decir la cantidad que la gente está dispuesta a comprar. Si el precio sube mucho, los consumidores se vuelcan a bienes alternativos, por ejemplo papas. Si el precio baja mucho comprarán más de lo estrictamente necesario sin preocuparse demasiado de lo que les sobre.
- Cuanto mayor es su precio, mayor será la oferta, es decir la cantidad que los panaderos están dispuestos a suministrar. Si el precio es lo suficientemente interesante trabajarán horas extras, usarán harina de sus stocks de reserva, etc. Si el precio es muy bajo concentrarán sus esfuerzos en producir factura y bizcochuelos, con poca producción de pan.

Podemos resumir lo dicho en un gráfico con el precio en el eje vertical y las cantidades demandadas y ofrecidas sobre el eje horizontal (por ahora mirar sólo las líneas llenas):

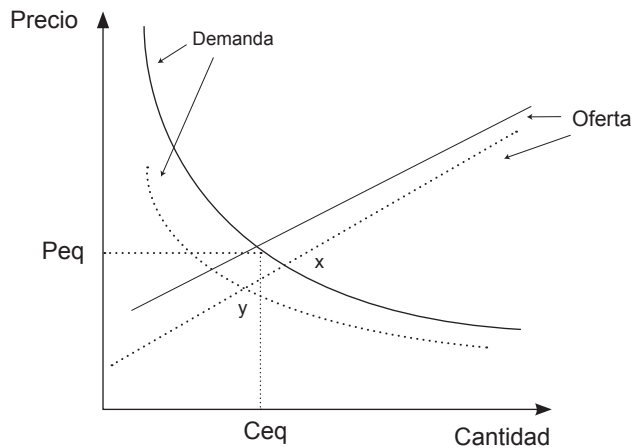

Figura 1.B.1

El precio y las cantidades en equilibrio vienen dados por la intersección de las curvas de oferta y demanda, es decir por los valores *Peq, Ceq*. A precios mayores los proveedores traen más pan al mercado del que el público desea adquirir a ese precio (oferta mayor que demanda), lo que los obliga a bajar el precio o quedarse con pan "de clavo". A precios menores hay escasez (más compradores que pan disponible), lo que permite subir el precio sin perder ventas.

El argumento clásico dice que si la curva de oferta se desplaza hacia abajo como lo muestra la línea punteada (los proveedores bajan costos o reducen su margen de ganancia) aumentan las ventas. En efecto, el equilibrio se desplaza al punto x con mayor cantidad vendida. Esto es indudablemente válido *siempre y cuando la curva de demanda no se mueva.* El economista lo expresa diciendo que *una baja de la curva de oferta **ceteris paribus**[15] aumenta la cantidad vendida.*

Si las restantes cosas no permanecen iguales el argumento se derrumba. Supongamos por ejemplo que, simultáneamente con la baja de precios de los panaderos, bajó muchísimo la papa. La demanda de pan puede haberse desplazado como indica la otra línea punteada, con lo que el equilibrio pasa al punto *y* con ventas menores, a pesar de haber bajado el pan. La condición ceteris paribus hace que el argumento sea esencialmente microeconómico, para mercados lo suficientemente pequeños como para no afectar al resto de la economía.

---

15. Ceteris paribus = "mientras las restantes cosas permanecen iguales" en latín

Si el "bien" ofrecido es la totalidad de la mano de obra del país en lugar de pan, el argumento pierde toda validez. Cualquier cambio del precio de oferta modifica la capacidad de consumo de la gente, por lo tanto la demanda de todos los otros bienes, y por lo tanto la demanda de mano de obra. ¿Ceteris paribus? De ninguna manera!!

Y aparte, la mano de obra tiene características específicas que la diferencian de otros bienes. Suponer que la oferta crece con el precio implica que los millones de argentinos con dos o tres trabajos renuncian al ocio y esparcimiento tentados por los excelentes salarios. Y que de recibir mayor paga, trabajarían aún más. Esto es un disparate total. Mucho más razonable es suponer que la oferta de mano de obra aumenta al bajar el precio, ya que hará falta trabajar más para subsistir. Aceptado esto, el problema cambia cualitativamente. Puede demostrarse que con curvas de oferta y demanda como muestra la figura 1.B.2 el equilibrio se vuelve inestable: por las fuerzas del mercado el sistema no va **hacia** el equilibrio sino que **se aleja** de él. No daremos la demostración, aunque invitamos al lector a intentarla (basta suponer hoy un precio ligeramente superior o inferior al de equilibrio y ver qué pasa en días sucesivos). Digamos por último que la inestabilidad mencionada corrobora lo hallado mediante un enfoque totalmente distinto en la sección 1.5.1.

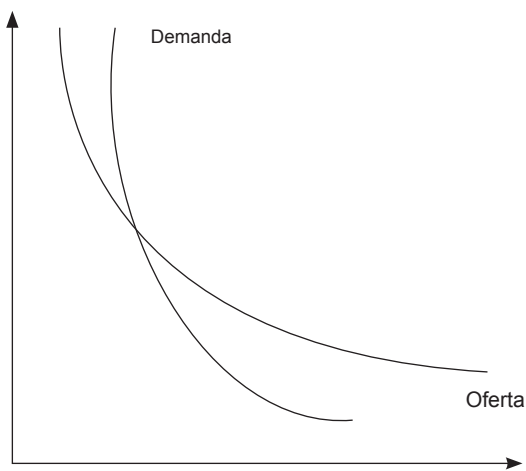

Figura 1.B.2

# 2. NOCIONES BÁSICAS DE MACROECONOMÍA

## 2.1 Nomenclatura

Se llama Macroeconomía a la rama de la Economía que estudia sistemas económicos "grandes" como un país o una región (por ejemplo el Mercosur). Se ocupa de los grades agregados: producción total, nivel de empleo, inflación, etc. Se llama *Microeconomía* al estudio de sistemas lo suficientemente chicos como para no alterar las variables macro: una empresa, una familia, el mercado de oleaginosas, etc. Nuestro estudio se centra en la Macroeconomía.

Se llama *agentes económicos* a las personas físicas o jurídicas con actividad económica (trabajan, compran, venden toman préstamos, etc.). Según el tipo de actividad vamos a dividir a los agentes económicos en cuatro grandes *sectores*:

- Empresas (productoras)
- Particulares (consumidores)
- Gobierno (consumidor, y productor de servicios como educación, justicia, defensa, etc.)
- Sector externo (productor y consumidor)

¿Dónde ponemos a los autónomos, quiosqueros, profesionales, etc. que son particulares y a la vez producen servicios? Solución: desdoblamos las funciones, por ejemplo: Juan Pérez (particular, consumidor) y Consultorio Dr. Juan Pérez (empresa, produce servicios de salud). Similarmente con los dueños de empresas: José Tuerca (particular=, Rectificaciones "La Tuerca" (empresa).

Se llama *corto plazo* a un tiempo insuficiente para que cambien los métodos de producción (equipamiento, composición de la fuerza laboral, tecnología). Típicamente es de semanas o unos pocos meses. Se lama mediano plazo a un tiempo suficiente como para tener cambios moderados de dichas magnitudes, por ejemplo re-entrenamiento de la mano de obra para nuevas funciones. Típicamente es de uno a cinco años. Se llama largo plazo a tiempos mayores, que permiten cambios demográficos, acumulación de capital, nuevas tecnologías, etc. Esta clasificación es por supuesto bastante imprecisa, pero resulta útil para expresarse. Por ejemplo si digo "en el corto plazo una mayor producción requiere el empleo de más mano de obra" la frase se entiende.

## 2.2 El Ingreso Nacional o Producto Bruto Interno

Se llama *Ingreso Nacional Y, o Producto Bruto Interno PBI* al valor total de lo que produce un país en cierto período, contando tanto los bienes (autos, tomates, casa, etc.) como los servicios (atención médica, enseñanza, espectáculos, etc.) Ejemplo: el *PBI* argentino de 1995 fue de 281.000 millones de pesos. Quiere decir que harían falta 281.000 millones de pesos para comprar todo lo que se produjo en Argentina ese año.

Si bien la idea es simple, tanto la definición precisa como la medición del *PBI* son problemas contables altamente no triviales. Veamos un "botón de muestra". Un agricultor vende $150 de trigo a un molino, que lo transforma en $300 de harina vendida a un panadero. Este a su vez la usa para producir $500 de pan. ¿De cuánto fue la producción? La única respuesta sensata útil es: el valor final del pan, $500. No se le deben sumar el trigo y la harina porque ya están *incluidos* en los $500. Perfecto, ¿excluimos entonces toda producción de trigo y harina del cómputo del *PBI*? No, esto tampoco serviría: dejaríamos fuera del *PBI* la harina producida y vendida directamente a amas de casa en vez de panaderos. Y nuestro propósito es contar todo lo que se produce una y sólo una vez, sin saltear y sin repetir.

La solución es computar como producción de cada agente económico el *valor agregado* que genera: su producto menos los insumos usados. Con este criterio tenemos para el panadero un valor agregado de $500-$300 = "200, para el molino $300-$150 =$150, y para el agricultor $150. La suma da los $500 como debe ser. Si el agricultor a su vez hubiera comprado $50 de semillas y fertilizantes, su valor agregado sería sólo $100, pero los $50 restantes aparecerían como valor agregado de los proveedores de semillas y fertilizantes.

En definitiva el uso de valores agregados es equivalente a calcular el *PBI* en base a los *productos finales* producidos. Los $500 de pan incluyen automáticamente los bienes intermedios trigo y harina; una heladera de $1000 incluye automáticamente los cables, chapa y motor incorporados a la misma, etc. Pero el concepto de valor agregado es más flexible: mientras la harina no se vendió ¿es producto final o no? No lo sabemos, depende de quién la vaya a comprar. La noción de valor agregado, al no distinguir entre bienes finales e intermedios, no nos plantea tal problema.

Vamos a establecer ahora la ecuación básica que relaciona el *PBI* con la actividad en los diversos sectores de la economía. Damos aquí una justificación heurística, y en el apéndice 2. A una deducción rigurosa. Consideremos todos los bienes y servicios producidos en el año. Según su destino se pueden clasificar en:

- *C*: lo adquirido por particulares para su consumo.
- *G*: lo adquirido por el Gobierno para su consumo.
- *Iv*: lo adquirido por alguien como inversión. Maquinaria, edificios, vehículos, capacitación de personal, acopio de materiales para períodos futuros, etc. por empresas. Viviendas por particulares. Grandes obras públicas por el gobierno.
- *Xn*: lo exportado (exportaciones netas, es decir exportaciones menos importaciones),
- *Q*: lo que queda sin vender[1]. Es sobreproducción respecto a las ventas. Un valor negativo de *Q* significa sub producción (venta de stocks previamente acumulados).

Como cada bien o servicio cae en alguno de estos grupos tenemos la siguiente igualdad, que juega un rol fundamental en la teoría macroeconómica

$$Y = PBI = C + G + Iv + Xn + Q \qquad (2.1)$$

Los diversos bienes y servicios se evalúan en la práctica a precios de mercado. Cuando nos interese comparar producciones físicas (estudiar el crecimiento del *PBI* en el tiempo, o comparar diferentes países) deberemos deflacionar los valores con un índice de precios y expresarlos todo en *moneda constante*, por ejemplo en u$s de 1990.

El cálculo hecho ignora las pérdidas por desgaste de maquinarias durante el proceso productivo. Ignora igualmente las pérdidas de valor de los equipos por obsolescencia. Llamemos *D* (depreciación) a la suma de tales pérdidas. Se llama producto Neto Interno *PNI* al *PBI* corregido por depreciación: *PNI = PBI – D*. En la práctica es casi imposible medir en forma medianamente confiable la depreciación, había que contestarse preguntas como ¿cuántos años servirá el camión comprado hoy? ¿cómo prorratear su costo a lo largo de su vida útil?, etc. Por tales dudas el concepto de *PNI* es poco usado, las cifras oficiales en particular se limitan al *PBI*. A "ojo de buen cubero" la depreciación se estima en 10% del *PBI* para economías industriales.

---

[1]. Los economistas acostumbran llamar a nuestro Iv **inversión ex-ante** o **inversión voluntaria**, a nuestro Q **inversión involuntaria**, y a nuestro Iv + Q, **inversión ex-post**. Para aumentar la confusión, muchos autores usan el mismo símbolo I para las inversiones ex-ante Iy ex-post. Tal nomenclatura no resulta conveniente para nuestros propósitos.

El apéndice 2.B da algunas pistas útiles para interpretar las cifras del Ingreso Nacional.

## 2.3. Sobre la igualdad entre Producción e Ingreso

La producción es, si se quiere, el *tamaño de la torta* que hemos producido entre todos. El *ingreso* de un agente económico (un particular, una empresa, un municipio o lo que sea) se define como la parte del *PBI* que le toca en el reparto. Es, si se quiere, *la tajada con que se queda cada uno*. Por tratarse de un reparto, está claro que la suma de los ingresos de todos los agentes económicos iguala a la producción. Si la repartimos íntegramente, la suma de las tajadas es igual a la torta original.

Mirémoslo de otra manera. Un agente económico puede tener ingresos de dos formas: a) porque produjo algo, o b) a expensas de otro agente económico (un buen ejemplo es: le ganó al póker). Al sumar sobre todos los agentes, los ingresos del tipo b) se cancelan: por cada ganador hay algún perdedor con ingreso equivalente negativo. La suma se reduce entonces a los ingresos originados en producción, y tenemos nuevamente (Suma de todos los ingresos) = Producción.

Los ingresos del tipo b), en que alguien recibe a expensas de otro, se llaman *ingresos de transferencia*. Ejemplos más serios son las jubilaciones, pensiones y subsidios (ingresos negativos para el Estado, positivos para algún otro), las donaciones, etc.

Los textos elementales identifican el ingreso de un agente económico con "lo que gana". Esto no es exacto. Ejemplo: una empresa produce 1.000 heladeras de $1.000, paga $600.000 por insumos y $200.000 de sueldos. Llega fin de año y sólo vendió 800. ¿Cuánto ganó en el año? Nada, por ahora "va hecha". ¿Cuál fue su ingreso en el año? $200.000, se quedó con una "tajada" de 200 heladeras.

En el apéndice 2.C se discute la discriminación del ingreso por sectores.

## 2.4 Producción de equilibrio

Las empresas tratan de mantener sus stocks en ciertos niveles que consideran óptimos. Si un mes hay sobreproducción *(Q > 0)* producen menos el mes siguiente para no acumular stocks inútiles. Si $Q < 0$ producen más para no perder ventas por falta de stock. En ambos casos la economía evoluciona hacia $Q = 0$, situación que por ese motivo se llama de equilibrio. La ecuación de equilibrio es la (2.1) con $Q = 0$

$$Y = C + G + Iv + Xn \qquad (2.2)$$

Se llama demanda efectiva al lado derecho de la ecuación (2.2): *C + G + Iv + Xn* Representa cuánto de la producción encuentra comprador, según se ve de las definiciones de *C, G, Iv* y *Xn* dadas en la sección 2.2

## 2.5 Estimaciones numéricas

En base a una publicación del Ministerio de Economía (Proyecciones Macroeconómicas 1995-99), a información periodística, y a consultas personales con especialistas, he recopilado valores aproximados para la economía argentina durante 1995 en la tabla 2.1 (montos expresados en millones de pesos). El lector que disponga de datos más actuales y/o confiables puede sustituirlos en las discusiones siguientes; las conclusiones no variarán mayormente.

| Cantidad | % del *PBI* | Monto |
|---|---:|---:|
| *PBI* | 100% | 281.000 |
| Consumo particulares *C* | 69% | 194.000 |
| Gastos Gobierno *G* | 13% | 36.700 |
| Inversión + acum. stocks *Iv.+Q* | 18% | 49.900 |
| Exportaciones netas *Xn* | aprox. 0% | 400 |
|    Exportaciones *X* | 8% | 23.800 |
|    Menos Importaciones *M* | -8% | -23.400 |

Tabla 2.1

## 2.6 Ingreso Nacional y nivel de empleo

En el corto plazo la tecnología y composición de la fuerza laboral son datos fijos, no hay tiempo para robotizar las fábricas o formar el cadete como analista de sistemas. En consecuencia las variaciones de producción requieren variaciones acordes en el nivel de empleo, ya sea mediante horas extras o la contratación de más gente. La relación entre *PBI* y nivel de empleo tiene la forma esquematizada en la figura 1.2, en la que ahora hemos dado un significado preciso al término "Producción": se refiere al *PBI*.

En la Argentina el desempleo ronda el 20%, quiere decir que *N* es aproximadamente un 80% del nivel de pleno empleo *NP*. Para solucionarlo, *el PBI debería aumentar grosso modo un 20%*, es decir en 56000 M$[2]. Con tal aumento el nivel de empleo pasaría del 80% actual a un aceptable 96% (80 más el 20% de 80 da 96).

## 2.7 El equilibrio Keynesiano – Función de consumo.

Del *PBI* los particulares reciben una cierta fracción *fp < 1* en sueldos y beneficios de las empresas (recordar: los accionistas y dueños de empresas son particulares). Si por ejemplo el 70% del *PBI* va a los particulares diremos que *fp = 0,7*. El ingreso de los particulares es *Y(P) = fp * Y*. De este ingreso consumen una fracción *cp*, y ahorran el resto. El consumo es entonces *C = cp * Y (P) = cp * fp * Y*. Llamando *c = cp * fp* (c minúscula para distinguirla del consumo) tenemos la función de consumo keynesiana [3,4]

$$C = c * Y \qquad (2.3)$$

El numerito *c* se llama *propensión marginal al consumo*. Por ser producto de dos valores positivos y menores que 1, resulta *0 < c < 1*. El valor de c cambia de un país a otro y de una época a otra pero puede considerarse fijo para un país en el corto plazo.

Cabe acotar que la función de consumo es un concepto estadístico. Es imposible prever cuándo un automovilista tendrá un accidente. Pero si hablamos de cientos de miles de individuos (que es el caso tratado por las compañías de seguros) rigen leyes probabilísticas que permiten predecir el número de siniestros con gran exactitud. Del mismo modo, la función de consumo (2.3) no pretende describir el consumo de un individuo, sino el comportamiento global de una gran población.

---

2. 1 M$ = $1.000.000
3. Propuesta por Keynes en 1936 para explicar la recesión de la década del ´30. Nuevamente hemos simplificado la matemática para hacer el tema accesible a un público lo más amplio posible.
4. Por razones no del todo claras, hay autores que consideran la (2.3) una funcion de consumo "no keynesiana", por ejemplo Enrique Ballesteros, *Introducción a la Teoría Económica*, Alianza Universidad Textos, Madrid 1988, pág. 433.

Sustituyendo el *C* de la (2.3) en la ecuación de equilibrio (2.2) obtenemos

$$Y = c * Y + Iv + G + Xn \qquad (2.4)$$

que permite despejar Y. Pasando el término c * Y al lado izquierdo y sacando Y de factor común resulta *(1 – c) Y = Iv + G + Xn* de donde

$$Y = (Iv + G + Xn) / (1 – c) \qquad (2.5)$$

e Introduciendo la cantidad *m = 1/(1 – c)* llamada *multiplicador keynesiano* la ecuación anterior queda

$$Y = m * (Iv + G + Xn) \qquad (2.6)$$

Dados *Iv, G y Xn* y los hábitos de consumo de la sociedad (reflejados en *m*), la (2.6) determina el nivel de producción total *Y*.

Si *Iv* aumenta en una unidad (digamos un millón de $) el paréntesis de la (2.6) aumenta en una unidad y por lo tanto *Y* aumenta en *m* unidades. Esto se conoce como *efecto multiplicador* de las inversiones: un aumento de *Iv* produce un aumento varias veces mayor de la producción total *Y*. Recíprocamente una caída de *Iv, G o Xn* produce una caída *m* veces mayor de la producción.

Conclusión: para actuar sobre *Y* el Gobierno dispone de al menos dos herramientas:

- Puede implementar medidas fiscales y/o monetarias que afecten *Iv o Xn*.
- Puede variar sus propios gastos *G*.

Los grandes méritos de la teoría keynesiana son:

- Explica por qué una economía puede alcanzar un equilibrio indeseable con alto desempleo. La ecuación (2.6) no contempla explícitamente el tamaño de la fuerza laboral. Sería por lo tanto una *enorme coincidencia* que el *Y* resultante sea el de pleno empleo.
- Sugiere posibles acciones correctivas.

## 2.8 Versión rigurosa del argumento sección 1.6.3

En la sección 1.6.3 hicimos un gráfico (figura 1.3) con "Producción" y "Ventas", dos términos que no estaban definidos. Cometimos el típico pecado del "guitarrero": usamos palabras sugestivas sin darles un significado preciso. De haber querido calcular algo en ese momento nos hubiéramos encontrado en dificultades (por ejemplo: la harina que el molino le vende al panadero ¿Se cuenta como producción? ¿O sólo va el pan?) Estamos ahora en condiciones de redimirno y dar una versión "prolija" de aquel argumento. Reemplazando el *C* de la ecuación (2.3) en la (2.1) tenemos $Y = c * Y + G + Iv + Xn + Q$, de donde se despeja

$$Q = Y - (c * Y + Iv + G + Xn) \qquad (2.7)$$

Construyamos un gráfico con *Y* en el eje horizontal, y representemos en él las dos funciones *PBI = Y y Demanda Efectiva = c * Y + Iv + G + Xn*. Según (2.7) la diferencia entre estas dos funciones da la sobreproducción *Q*. La primera función *PBI = Y* es una recta de pendiente 1, es decir a 45º con los ejes. La segunda es otra recta, que arranca de *Iv + G + Xn para Y = 0,* y tiene pendiente *c < 1*. Los gráficos son los que muestra la figura 2.1

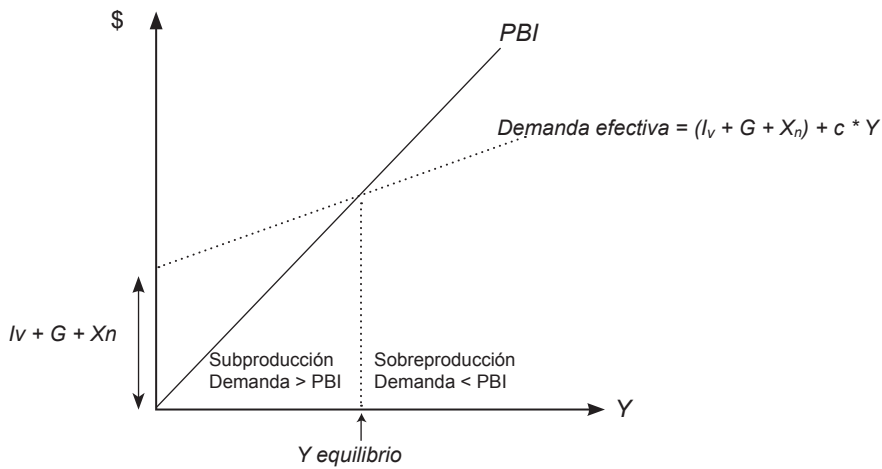

Figura 2.1

La ecuación 2.7 y su representación gráfica figura 2.1 son la versión prolija buscada. Los vocablos ambiguos han sido sustituidos por el *PBI* y la *deman-* *da* efectiva, de los cuales **sabemos qué son y cómo se calculan**. Sobre el eje horizontal ahora tenemos *Y* en lugar del nivel de empleo *N*. Esto es irrelevante, se trata de un simple cambio de escala, como si en vez de medir longitudes en cm lo hiciéramos en pulgadas. En efecto, estamos suponiendo *N* e *Y* proporcionales (ver sección 2.6).

A partir de aquí el argumento de 1.6.3 sigue intacto: la intersección da el punto de equilibrio, una baja del salario produce una línea de demanda más horizontal, etc. llegando a la misma conclusión: la baja del salario real *disminuye* la producción. Dejamos los detalles para el lector

Al presentar esta discusión a oyentes con formación macroeconómica, se me ha objetado: 1) que las diversas funciones que aproximo por rectas en realidad no son tales, y 2) que al bajar el salario real aumenta *N*, con lo cual el consumo total puede aumentar en lugar de disminuir. Contesto a ambas en el apéndice 2.D. Dejamos aquí este tema, y retomamos la teoría del multiplicador y sus enseñanzas.

## 2.9 Evaluación de posibles medidas económicas.

Comencemos por conocer el multiplicador para el caso argentino: Según la Tabla 2.1, *c = 0,69* para 1995. Por lo tanto

$$m = 1 / (1 - c) = 1/0,31 = 3,23 \qquad (2.8)$$

### 2.9.1 Exportaciones

Una primera observación es que todos los países tienen problemas de desempleo. Brasil (por dar un ejemplo) nos comprará en la medida en que hay reciprocidad. Evidentemente vender y comprar por valores comparables no soluciona nada, ya que da *Xn = 0*. Argentina no tiene ventajas comparativas evidentes (como bajos costos, monopolio de algún producto, etc.) para ser exportador neto. Sin embargo, seamos optimistas. Supongamos que sea posible aumentar las exportaciones en un 100% en uno o dos años, mientras las importaciones crecen sólo un 50% (más hipótesis de trabajo es una expresión de deseos). Esto llevaría las exportaciones a 47.000 M$, las importaciones a 35.000 M$ y *Xn* a 12.000 M$. Usando el multiplicador (2.8) resulta

un aumento del *PBI* de 38.800 M$, o sea 69% del necesario para corregir la desocupación.

Evidentemente las medidas para aumentar Xn son bienvenidas, van en la dirección correcta. Y la apertura comercial aporta otras ventajas al consumidor. Pero aún con mucho optimismo las exportaciones sólo llegan a *reducir* el problema, no a *resolverlo*.

### 2.9.2 Inversión

Es un hecho que el país tiene capacidad ociosa. Habiendo máquinas paradas ¿para qué comprar otras? Inducir a los empresarios a aumentar *Iv* va a ser difícil. Por otro lado venimos suponiendo inversiones que *no cambian la tecnología existente*. Si en *Iv* hay otro tipo de inversiones (fábricas automatizadas, hipermercados, etc.) hay que revisar la teoría. La (2.6) sigue valiendo, pero la recta de Fig. 1.2 se hace más empinada en el transcurso del tiempo por mayor productividad. La inversión tiene entonces dos efectos opuestos: crea mayor *PBI* lo que tiende a aumentar *N*, e incrementa el rendimiento por hora-hombre lo que tiende a emplear menos gente. No está claro cual de los dos efectos domina.

### 2.9.3 Gasto gubernamental G

La tradicional "receta" keynesiana para combatir la recesión es aumentar el gasto público *G*, financiando el déficit con deuda. El uso continuo de esta política lleva a largo plazo a una deuda pública impagable, lo que efectivamente ya sucede para muchos países. En el caso concreto argentino los ingresos del Estado no le permiten ni siquiera pagar los intereses de la deuda[5], mucho menos el capital. Creo que sería irresponsable seguir apelando a este recurso. El *diagnóstico* keynesiano me parece impecable, el *remedio* tan malo como la enfermedad.

Se puede mejorar la "calidad" del gasto.

### 2.9.4 ¿Qué nos queda?

Aparentemente una sola alternativa: actuar sobre el consumo C. Volveremos sobre el tema en la parte 3.

---

5. Si vendo mi piano para pagarle intereses al banco no los estoy pagando con mis ingresos sino a expensas de mi activo. De la misma manera, la venta de las empresas públicas no es contablemente un ingreso.

## 2.10 Ahorro, consumo y política fiscal

Se define *ahorro* de un agente económico, *S(a)*, como la diferencia entre sus ingresos y su consumo: *S(a) = Y(a) – C(a)*. El a puede ser cualquiera: un particular, una empresa, un municipio, etc. Sumando sobre todos los agentes económicos tenemos el ahorro nacional *S = Y - (consumo total)*. El consumo consta del consumo privado *C* y del consumo gubernamental *G*[6]. En consecuencia

$$S = Y - C - G \qquad (2,9)$$

Reemplazando aquí el Y de la ecuación 2.1 se obtiene

$$S = Iv + Xn + Q \qquad (2.10)$$

Algunas breves observaciones sobre esta fórmula:

- Si suponemos *Q = 0* (equilibrio) y *Xn = 0* (no hay comercio exterior), recapturamos la igualdad *ahorro = inversión*, ya vista para la "economía de isla" en la sección 1.5.2.

- El tratamiento Standard llama a *Q inversión en stock* y define inversión total como *I = Iv + Q*. Haciendo esto la ecuación queda *S = I + Xn*, válida con o sin equilibrio (para cualquier *Q*).

- Hay autores que van más lejos y llaman a *Xn inversión externa*, con inversión total *I = Iv + Xn + Q*. Entonces queda simplemente *S = I*, *ahorro = inversión*, válida siempre como consecuencia de estas particulares definiciones.

Supongamos que un cierto día los particulares deciden ahorrar más, reduciendo su propensión al consumo del actual *c = 0,69 a c= 0,66*. El multiplicador pasa de *3,23 a 2,94*, lo que representa una caída del 9%. A igual *Iv, G y Xn* la fórmula (2.6) nos dice que *Y* se reduce asimismo en *9%* (recuperamos el resultado ya visto en la "isla"; el ahorro hace caer la producción).

Recíprocamente, si deciden consumir más llevando *c a 0,72*, el multiplicador pasa de *3,23* a *3,57*, con un aumento porcentual del 10% en *m* y en *Y*.

---

6. Según el apéndice 2.1 el gasto gubernamental es *G= C(G) + W (G)* donde *C(G)* representa cosas compradas a empresas, y *W(G)* el valor de los servicios provistos por el Gobierno. Es una cuestión de gustos clasificar los últimos como consumo del gobierno, o de los particulares que son sus beneficiarios. El hecho es que tales servicios no dejan bienes durables y por lo tanto son consumo.

Conclusión: cuando hay recesión **el ahorro es el villano de la película**. La contracara del ahorro, el *consumo*, es "el muchachito" que puede salvarnos.

El impuesto más importante en la Argentina es el IVA. Lo paga quien consume algo, y no lo paga quien ahorra. Es decir que pone premios y castigos exactamente en los lugares equivocados si se quiere reactivar.

Hasta aquí siguiendo a los libros de texto, supusimos que *c* lo determinan los particulares con sus hábitos de ahorro y consumo. En realidad no es así. Una buena parte del ahorro del país lo realizan las propias empresas para financiar sus operaciones. Los datos se pueden ver en los balances publicados: siempre hay un ítem "beneficios no distribuidos" que en general supera ampliamente lo pagado en dividendos.

¿Qué sucede si las empresas deciden ahorrar más? Baja el ingreso de los particulares (porque baja el *fp* de la sección 2.7) y por lo tanto la propensión marginal al consumo *c = cp \* fp*. Baja asimismo m = *1/ (1-c)*. A partir de allí pueden suceder dos cosas.

- O bien el ahorro de las empresas se vuelca a la compra de equipamiento (inversión). En tal caso no hay ningún problema. En la ecuación 2.6 disminuye *m* pero aumenta *Iv*, siendo el fecto neto incluso *reactivador*. Para ver esto último conviene volver a la ecuación 2.1. Si la empresa ahorra $30.000 y compra un camión, *Iv* aumenta en $30.000. El ingreso de los particulares disminuye en $30.000, de los cuales habrían consumido una fracción *cp*. Luego, el consumo cae en $30.000 \* *cp*. La variación total de *Y* es *$30.000 \* (1-cp) > 0* por ser *cp < 1.*
- O bien el ahorro se dedica a bicicletas financieras, a la adquisición de otras empresa, etc. que no implican la adquisición de producción nacional. Entonces queda sólo la caída del consumo por los particulares. El efecto es recesivo, como si ahorraran éstos.

Lo deseable sería una política fiscal que vigile de cerca estos aspectos y aplique las presiones del caso.

**2.10.1 Disgresión: qué se consiguió con la deuda externa.**

Si Juan ahorra $1.000, baja el consumo y produce desempleo. Si al mismo tiempo Pedro saca un crédito por $1.000 y compra un lavarropas, cancela el ahorro de Juan. Ha restablecido el consumo original. El ahorrista renuncia a consumo presente y lo posterga para el futuro, el tomador de crédito hace exactamente lo opuesto: consume hoy a cuenta de ingresos futuros. Si por cada individuo que ahorra consiguiéramos uno dispuesto a endeudarse, no habría problema de desempleo. En la práctica, parte del ahorro privado es tomado en préstamo por los empresarios para sus inversiones, a través del sistema bancario. Pero aún así sobra ahorro, y harían falta algunos "perejiles" más, dispuestos a endeudarse para absorberlo.

Para lo anterior no importa si Pedro reside en el país o en el exterior, mientras tome el crédito y lo gaste aquí va todo bien. Pedro podría incluso ser un gobierno extranjero.

A la luz de estas consideraciones está claro que en los últimos 20 años las economías industrializadas **resolvieron buena parte de sus tensiones sociales y problemas de desempleo gracias a los países periféricos.** Estos últimos fueron (fuimos) los "perejiles" que acumularon deudas impagables para consumir en muchos casos baratijas, impulsando así economías ajenas y deprimiendo las propias.

## 2.11 Limitaciones de la teoría.

La propensión marginal al consumo *c* depende no solo de las variables económicas sino también de factores psicológicos de los consumidores: expectativas de ingresos futuros, optimismo o no respecto de las medidas gubernamentales, situación mundial, etc., todos factores cambiantes con el tiempo. Como ya se mencionó en 2.7, *c* varía de un país a otro y de una época a otra. En consecuencia el análisis hecho desde la sección 2.7 en adelante sólo **es *útil* en el corto plazo.**

Para predicciones a mediano y largo plazo no conocemos los valores futuros de *c* de modo que, aunque las diversas fórmulas sigan valiendo, no tienen capacidad predictiva. Nos encontramos en la situación de quien quiere medir logintudes con una regla de "chicle". Esto será tenido en cuenta al formular propuestas en la parte 3 del presente trabajo.

## Apéndice 2.A – Cómputo del PBI

El Producto Bruto Interno *PBI* pretende medir mediante un número el total de la producción de un país en cierto período. La disciplina que estudia su cómputo es la *Contabilidad Nacional*, para cuyo dominio se requiere conocer una enorme cantidad de tecnicismos que superan holgadamente los propósitos de este trabajo. Trataremos de exponer en este apéndice solo los aspectos más fundamentales del tema.

La producción consta de cosas heterogéneas: casas, autos, arvejas, etc. Para sumarlas hay que llevarlas a una unidad común, que es su *valor en dinero:* $1.000.000 en ladrillos y $50.000 en lapiceras dan $1.050.000 de producción. El *PBI* se mide en *dinero.*

¿Qué se considera producción? Todo lo creado en el año que tenga *valor.* Y el valor de una cosa es por definición lo que la gente está dispuesta a pagar por ella: el valor se mide por el *precio de mercado*. Con esta concepción se incluyen tanto bienes como servicios. La gente compra nafta y clases de piano, luego ambos ítems forman parte del *PBI*.

Al definir los sectores de la economía hemos acordado que los particulares son sólo consumidores, de modo que podemos escribir.

$$PBI = P(E) + P(G) \qquad (2.A.1)$$

donde *P(E)* es la producción de las empresas y *P(G)* la producción del gobierno. Lo producido por el sector externo obviamente no va, ya que queremos evaluar la producción de *nuestro* país, y lo producido en Brasil forma parte del *PBI* brasileño, no del nuestro.

Consideremos una empresa individual y su producción $P(e)$[7]. Tenemos

$$P(e) = ValProd(e) - Insumos(e) \qquad (2.A.2)$$

donde *ValProd(e*) es el valor de los bienes producidos, e *Insumos(e)* es el valor de los insumos comprados a terceros. Por ejemplo, si un panadero compra harina por $300 y con ella produce pan por $500, el valor de su producción (su trabajo de transformar la harina en pan y comercializarlo) es de $200. En otras palabras, identificamos *P(e)* con el *valor agregado* que genera la empresa.

*ValProd(e)* será en general casi igual a las ventas del período *Ven(e)*, pero la coincidencia no es exacta: pueden venderse hoy artículos en stock desde el año pasado, o pueden producirse hoy artículos que recién se venderán el año próximo. La expresión correcta es

$$ValProd(e) = Ven(e) + VarStk(e) \qquad (2.A.3)$$

---

7. Notación: usamos e minúscula: *P(e)* para referirnos a una empresa individual, y mayúscula: *P(E)* para la totalidad de las empresas, o sea todo el sector. La misma convención se usará en todos los casos.

Donde *Var Stk(e)* es la variación de stocks entre el comienzo y el fin del período, contada como positiva si el stock aumentó y negativa si disminuyó.

Por un argumento similar, *Insumos(e)* será aproximadamente igual a *Comp(e)*, las compras del período, con la salvedad de que las compras pueden incluir equipamiento o acopio de materiales, que no constituyen insumos sino inversión[8]. La fórmula correcta es

$$Insumos(e) = Comp(e) - I(e) \qquad (2.A.4)$$

Siendo *I(e)* la inversión. Sustituyendo (2.A.3) y (2.A.4) en (2.A.2) obtenemos

$$P(e) = Ven(e) - Comp(e) + Var\ Stk(e) + I(e) \qquad (2.A.5)$$

Tenemos ahora P*(e)* expresado en términos de cantidades accesibles a la Contabilidad Nacional. Las compras y ventas son conocidas a través de las declaraciones de IVA, los otros dos datos de los balances de la empresa.

Discriminemos ahora las ventas según el sector al que pertenece el comprador:

$$Ven(e) = Ven(e,P) + Ven(e,E) + Ven(e,G) + Ven(e,X) \qquad (2.A.6)$$

donde los cuatro términos del lado derecho son las ventas de la empresa a particulares, a otras empresas, al gobierno y al exterior respectivamente. Hagamos lo propio con las compras, teniendo en cuenta que por definición ni particulares ni gobierno venden cosas:

$$Comp(e) = Comp(e,E) + Comp(e,X) \qquad (2.A.7)$$

Sustituyendo (2.A.6) y (2.A.7) en (2.A.5) obtenemos la intimidante fórmula

$$P(e) = Ven(e,P) + Ven(e,E) + Ven(e,G) + Ven(e,X) +$$
$$- Comp(e,E) - Comp(e,X) + Var\ Stk(e) + I(e) \qquad (2.A.8)$$

---

8. El panadero puede haber comprado una balanza, o harina para el año próximo.

El paso final es sumar (2.A.8) para todas las empresas y así obtener la producción *P(E)* del sector:

$$P(E) = Ven(E,P) + Ven(E,E) + Ven(E,G) + Ven(E,X) +$$
$$- Comp(E,E) - Comp(E,X) + Var\,Stk(E) + I(E) \qquad (2.A.9)$$

En esta expresión los términos *Ven(E,E) y Comp(E,E)* se cancelan ya que una misma transacción entre empresas aparece como venta para una de ellas y compra para otra. Las ventas *Ven(E,P)* a particulares serán en general artículos de consumo, aunque puede también haber artículos durables que constituyan inversión. En consecuencia, *Ven(E,P) = (Consumo particulares) + (Inversión particulares) = C(P) + I(P).* Similarmente, *Ven(E,G) = C(G) + I(G).* Llamemos *X = Ven(E,X)* a las ventas al exterior (exportaciones), y *M = Comp(E,X)* a las compras al exterior (importaciones). Entonces *Ven(E,X) – Comp(E,X) = X – M = Xn* = exportaciones netas. Por último, abreviemos *VarStk(E) = Q*, son los artículos producidos que no se han vendido. Con todo esto la (2.A.9) toma la forma más explícita y manejable.

$$P(E) = C(P) + I(P) + C(G) + I(G) + Xn + Q + I(E) \qquad (2.A.10)$$

que completa el cálculo de la producción por parte de las empresas. Para el *PBI* nos resta evaluar el otro término de (2.A.1), a saber la producción del gobierno, lo que afortunadamente es mucho menos engorroso. *P(G)* consiste en los diversos servicios que el gobierno brinda: educación, salud, justicia, defensa, etc.

El público paga, digamos $20 la hora por una clase de japonés, la que entonces se computa en el PBI a ese valor. ¿Cuánto paga por la administración de justicia? Paga[9] los insumos de los juzgados (luz, gas, resmas de papel, etc.) más los sueldos del personal judicial (desde porteros hasta miembros de la Corte Suprema). Los insumos no son producción del gobierno sino de la empresas, y como tal aparecen incluidos en (2.A.10), concretamente en el término *C(G)*. Resta como producción del gobierno el gasto en personal. Por supuesto el mismo análisis vale para los otros servicios (educación, seguridad, salud, etc.) de modo que

$$P(G) = W(G) \qquad (2.A.11)$$

---

9. Por supuesto que en forma indirecta, a través de los impuestos votados (en teoría) por sus representantes en el Congreso.

*66    Coniglio*

donde *W(G)* es la totalidad de sueldos pagados por el gobierno. Con esto resulta

$$PBI = C(P) + I(P) + C(G) + I(G) + Xn + Q + I(E) + W(G) \quad (2.A.12)$$

Finalmente, llamemos *C* a secas al consume privado *C(P)*. Llamemos gastos del gobierno *G* a la suma de sus compras de insumos *C(G)* más su gasto en sueldos *W(G)*. Y llamemos inversión voluntaria total Iv a la suma *I(P) + I(G) +I(E)*. Con esta notación la (2.A.12) se convierte en la igualdad (2.1)

$$PBI = C + G + Iv + Xn + Q \quad (2.1)$$

Es de destacar que (2.1) no es un resultado de la teoría keynesiana, aunque se la use profusamente en la misma. Representa una identidad contable, **válida siempre por la definición de las cantidades intervinientes**, e independiente de cualquier teoría macroeconómica particular.

**Notas aclaratorias:**

- Puede haber empresas que aumenten su stock por decisión propia, no porque les quede producción "de clavo". Tales variaciones se deben tomas como *inversión en stock*, e incluirlas en *I(E)*, no en *Q.*
- Se considera a las empresas como exclusivamente productoras, sin consumo (no hay ningún *C(E)* en las ecuaciones). Todo lo que las empresas compran se computa o como inversión o como insumo. El café para los muchachos de la oficina es un insumo al igual que el acero si lo paga la empresa, o un consumo particular si se lo pagan ellos mismo.
- Sobre todo para los particulares, la distinción entre bienes durables (inversión) y consumo puede llegar a ser difusa. Si tomo clases de inglés para mi profesión ¿es consumo o inversión? ¿Y si tomo clases de guitarra por placer? Para no incurrir en ambigüedades, la Contabilidad Nacional corta por lo sano: define como consumo todas las compras de los particulares, excepto los inmuebles que considera inversión.
De todos modos incluir otros productos en la inversión (por ejemplo automóviles) no afectaría demasiado el análisis, simplemente pasaríamos cierto monto del término *C* al *Iv* de la ecuación (2.1), lo que no altera el

*PBI*. Por el mismo motivo, la definición exacta de la inversión pública *I(G)* no tiene mayores consecuencias, cambiándola se pasa cierto monto de *Iv* a *G* sin alterar el *PBI*.

## Apéndice 2.B – A tener en cuenta al interpretar el *PBI*

El *PBI per cápita* se obtiene dividiendo el *PBI* por el número de habitantes del país. Se lo llama también *renta per cápita* y pretende medir en cierta forma el nivel de bienestar del país. En el caso argentino su valor para 1995 es aproximadamente $281.000.000.000 / 33.000.000 = $8.515, o sea $710 mensuales por habitante, lo que no está nada mal. Por supuesto el PBI no da ninguna información sobre cómo se distribuye esta renta (que es un promedio) entre los diversos habitantes. Y está claro que en la Argentina muy pocas familias tipo (matrimonio + dos hijos) tienen ingresos de $2.800 mensuales, de modo que la distribución del ingreso es poco equitativa.

Para comprar *PBIs* de diferentes países se los convierte a la misma moneda, por ejemplo podrían compararse las rentas per cápita de Argentina y EE.UU. medidos en u$s. El uso de una moneda común no necesariamente elimina distorsiones, ya que nuestro PBI estará evaluado *a precios en u$s del mercado argentino*, mientras que el de ellos usa precios de mercado de allá, que pueden ser diferentes. Ejemplo: si llamo a mi cuñado a N. York y hablo 3 minutos, se calcula que Telefónica incorporó u$s 10 en servicios al PBI argentino. Si él me llama a mí, **el mismo servicio** se computa como u$s3 en el PBI estadounidense.

Igualmente, al comparar la producción en diversos períodos hay que tener sumo cuidado con las correcciones por inflación usadas. Digamos que queremos ver crecimiento del *PBI* argentino entre 1988 y 1995, reduciendo todo a u$s del año 1988 (lo que se llama base 1988=100). ¿Es correcto usar la pérdida de valor del u$s en el mundo (3 a 5% anual?) como factor de deflación? *De ninguna manera*, en ese lapso la Argentina tuvo una fuerte inflación en dólares por encima de la inflación mundial, que no estamos teniendo en cuenta. Digamos que tanto en 1988 como en 1995 se produjeron 1.000.000 de paquetes de cigarrillos (como fumador, recuerdo los precios de mercado: u$s 0.30 y u$s1.50 respectivamente). El aporte de esta producción al *PBI* es de u$s 300.000 en 1988 y de u$s 1.500.000 en 1995. Corrijamos el segundo valor por la caída del valor del dólar en 7 años, algo así como 30%, da $1.500.000/1,3 = $1.154.000. Seguimos teniendo una distorsión enorme, *la misma producción física aparece como un crecimiento del 284% en el PBI!!*

(Personalmente, creo que el espectacular crecimiento del PBI argentino que sale en los diarios se basa en alguna "manganeta" de este estilo).

Los libros de texto llaman al *Q* de nuestra ecuación (2.1) *inversión en stock* (es la producción que quedó sin vender). Y lo juntan con nuestro *Iv* bajo el nombre genérico de inversión: *I = Iv + Q*. Con esto la ecuación (2.1) queda *PBI = C + G + I + Xn*. Bien, supongamos que este año hubo un parate y quedó el 30% de la producción sin vender. Los economistas podrán hablar de un "sustancial aumento en las inversiones" **y técnicamente será correcto,** (Q aumentó!!) aunque el mensaje transmita un optimismo engañoso.

La medición del PBI dada por las cifras oficiales tiene un margen de error considerable, sobre todo en países con alta evasión fiscal. La Contabilidad Nacional no tiene más remedio que *estimar* la producción informal (vulgo: en negro), lo que distorsiona los resultados y da bastante libertad para "dibujar" los números.

En cualquier país se efectúan muchísimos trabajos no rentados que el PBI por lo tanto no registra. Por ejemplo, el de las amas de casa, o del individuo que construye su propia casa "a pulmón". Eso conduce a una serie de paradojas. Si pinto mi casa y el vecino la suya, no contribuimos al *PBI*. En cambio si yo pinto la de él y él la mía, y nos cobramos mutuamente $1000 por el trabajo, habremos aumentado el *PBI* en $2000, sin ninguna diferencia real entre ambas situaciones. Es fácil encontrar otros ejemplos similares. Esto de ninguna manera invalida la utilidad del *PBI*, ya que se trata más bien de curiosidades con poca incidencia cuantitativa.

## Apéndice 2.C – Discriminación del ingreso por sectores

Sean *Y(E), Y(P), Y(G)* e *Y(X)* los ingresos de los sectores empresas, particulares, gobierno y externo respectivamente. Las empresas producen por valor *P(E)*, de lo cual pasan a los particulares *W(E)* en sueldos y *B(E)* en beneficios empresarios (dividendos en las SA, retiros patronales en empresas menores). Al gobierno le "ceden" *T(E)* en forma de impuestos. Puede haber otras transferencias que por su escasa magnitud o por comodidad de análisis no nos interesa explicitar, las englobamos en un término "comodín" *R(E)*, o sea "el resto de los ingresos". Si la empresa recibe un subsidio, este se computará como positivo en *R(E)*, si dona camisetas de fútbol al equipo de la fábrica se computa como negativo por ser un egreso. Con todo esto, la "tajada" que le queda a la empresa es:

$$Y(E) = P(E) - W(E) - B(E) - T(E) + R(E) \qquad 2.C.1.1)$$

El sector de particulares recibe *W(E)* + *B(E)* de las empresas, *W(G)* del gobierno en sueldos de empleados públicos y le devuelve al gobierno *T(P)* en forma de impuestos personales. Agregando otro comodín *R(P)* por otros conceptos, tenemos

$$Y(P) = W(E) + B(E) - T(P) + R(P) + W(G) \qquad (2.C.1.2)$$

El gobierno produce *P(G)* = *W(G)* (ver ecuaciones (2.A.1) y (2.A.11), recibe *T(E)* + *T(P)* en forma de impuestos, y paga *W(G)* en sueldos. Le queda como ingreso

$$Y(G) = T(E) + T(P) + R(G) - W(G) + P(G) \qquad (2.C.1.3)$$

Donde el término *W(G)* se cancela (aparece positivo como producción y negativo al pagar sueldos), y *R(G)* es como siempre todo el resto. Por último, el sector externo no tiene ingresos explícitos por el momento, quedando simplemente

$$Y(X) = R(X) \qquad (2.C.1.4)$$

Sumando las cuatro ecuaciones y teniendo en cuenta que *R(E)* + *R(P)* + *R(G)* + *R(X)* = *0* por ser los *R* transferencias, se obtiene para el ingreso total Y:

$$Y = Y(E) + Y(P) + Y(G) + Y(X) = P(E) + P(G) = PBI \qquad (2.C.2)$$

es decir, ingreso total = producción como debe ser.

A partir de estas fórmulas básicas, es posible pasar a discriminaciones más detalladas. Por ejemplo, si nos interesara explicitar el pago de jubilaciones y pensiones *J*, y de intereses *D* sobre la deuda externa por parte del gobierno, restaríamos *J* y *D* de *Y(G)*, y los sumaríamos en *Y(P)* e *Y(X)* respectivamente para obtener

$$Y(E) = P(E) - W(E) - B(E) - T(E) + R(E) \qquad (2.C.3.1)$$
$$Y(P) = W(E) + B(E) - T(P) + J + R(P) \qquad (2.C.3.2)$$

$$Y(G) = T(E) + T(P) - J - D + R(G) \qquad (2.C.3.3)$$

$$Y(X) = D + R(X) \qquad (2.C.3.4)$$

donde por supuesto los varios *R* ahora tienen significados distintos: ya no incluyen *J* ni *D*. Si nos interesaran los "royalties" *Z* que las empresas pagan por patentes extranjeras, restaríamos *Z* de *Y(E)* y lo sumaríamos a *Y(X)*. etc. Las alternativas son muchas, las transacciones explicitadas dependen de lo que queramos estudiar.

## Apéndice 2.D – Detalles técnicos

Este apéndice está dirigido exclusivamente al lector con formación macroeconómica. Vamos a mostrar que nuestra conclusión de que menor salario real implica menor empleo no es fruto de las aproximaciones realizadas, sino que sigue valiendo en el caso general.

Consideremos la función de producción correcta $Y = Y(N)$ teniendo en cuenta rendimientos decrecientes de la mano de obra. Será una función creciente y con derivada segunda negativa, con la forma indicada por la curva Y de la figura 2.D.1

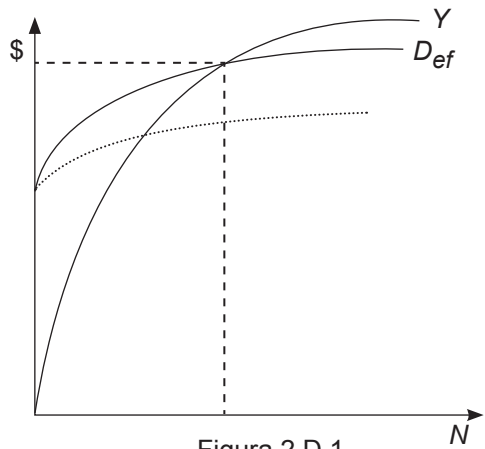

Figura 2.D.1

El consumo $C(Y)$ puede expresarse como función de función: $C = C(Y(N)) = C(N)$, lo mismo que la demanda efectiva $Def = C + G + Iv + Xn = Def(N)$. De vuelta sin suponer linealidad, tendremos alguna curva $D_{ef}(N)$ como se indica en la figura. Esto reproduce la figura (2.1) de la sección 2.8, con la sola diferencia de usar *N* como variable en las ordenadas. La intersección de *Y(N)* con *Def(N)* da el empleo de equilibrio como ordenada, y la producción como abscisa. Suponer ahora una baja del salario real. **Para cada *N*** la nueva demanda efectiva será inferior a la anterior, la nueva curva *Def(N)*, (indicada por la línea punteada), cae debajo de la previa. El punto de equilibrio se desplaza hacia abajo (menor producción) y hacia la izquierda (menor empleo). QED

# 3. PROPUESTAS CONCRETAS

En las partes 1 y 2 hemos dado sólidos argumentos teóricos para mostrar que **contrariamente a la creencia usual** el desempleo es causado por un salario real demasiado bajo. Hemos usado modelos simplificados (la "isla"), argumentos heurísticos (sección 1.6) y las ecuaciones standard de la macroeconomía (en la parte 2), arribando en todos los casos a la misma conclusión. Los datos empíricos (sección 1.6) confirman nuestra teoría y rebaten la creencia usual.

La solución, al menos en el corto plazo, parecería pasar entonces por un incremento del salario real. Esto merece algunas reflexiones. Para el hombre de la calle (y para más de un gurú televisivo) aumentar la masa salarial en, digamos $1.000.000.000, significa que otro sector, presumiblemente los empresarios, perderá igual monto. Esta opinión popular tan difundida, que voy a llamar *falacia de la torta fija*, es un caso especial de la falacia de composición. Si **un** empleador paga $10.000 más en sueldos, probablemente reducirá sus ganancias en esa suma. Pero si **todos** aumentan en forma proporcional se incrementa el consumo, con ello las ventas, y con éstas las ganancias. Es al menos *concebible* que las mayores ganancias compensen el gasto salarial extra. La torta **no** es fja. Y si crece, alguien *podría* llevarse una tajada más grande sin sacarle nada a nadie. Los párrafos siguientes exploran esta posibilidad.

### 3.1 Análisis a corto plazo

Vamos a usar las mismas ecuaciones macroeconómicas de la parte 2, peo redefiniendo los sectores en forma acorde a nuestro propósito presente. Dividamos los agentes económicos en tres sectores:

- Sector 1: los particulares de bajos ingresos (daremos una definición precisa más adelante)
- Sector 2: los restantes particulares (de ingresos medios y altos)
- Sector 3: Los otros agentes económicos (empresas + gobierno + sector externo)

Llamemos *f1, f2* y *f3* a la fracción del Ingreso Nacional que recibe cada sector. Tenemos entonces para el sector 1 un ingreso *Y1 = f1 \* Y*, y similar-mente para los sectores 2 y 3[1]. Por ejemplo si 1,2,3 reciben 30%, 20% y 50% del ingreso, será *f1 = 0,3, f2 = 0,2 y f3 = 0,5*. Obviamente

$$f1 + f2 + f3 = 1 \tag{3.1}$$

Para que la suma de los ingresos dé el total. Sean *c1* y *c2* las **fracciones** de sus ingresos que los sectores 1 y 2 consumen. Los consumos son entonces *C1 = c1 \* Y1 = c1 \* f1 \* Y y C2 = c2 \* Y2 = c2 \* f2 \** Y. El consumo combinado de ambos es

$$C = C1 + C2 = (c1 * f1 + c2 * f2) * Y \tag{3.2}$$

Y como en general es *C = c \* Y* resulta una propensión marginal al consumo

$$c = c1 * f1 + c2 * f2 \tag{3.3}$$

Supongamos que todos los valores anteriores describen un estado inicial de equilibrio con desempleo. Imaginemos ahora una redistribución del ingreso, consistente en reducir la fracción asignada a los sectores 2 y 3 en un porcentaje *p* (aclaración: no estamos reduciendo sus ingresos en *p* porciento, ya que como veremos la torta va a crecer). Usaremos para el nuevo equilibrio las mismas variables, pero distinguiéndolas con "primas", *v.g.f1', Y'*, etc. Tenemos entonces

$$f2' = f2 * (1-p) \tag{3.4.1}$$
$$f3' = f3 * (1-p) \tag{3.4.2}$$

El valor restante *f1'* sale de la (3.1): *f1' = 1 – (f2' + f3') = 1 – (f2 + f3) \* (1 – p) = 1 – (1 – f1) \* (1 –p)* donde en el último paso usamos nuevamente (3.1) para poner *f2 + f3 = 1 – f1.* Desarrollando el producto y simplificando

---

$$f1' = f1 * (1 - p) + p \qquad (3.4.3)$$

Con los mismos hábitos de consumo que antes (iguales *c1 y c2*), veamos qué propensión marginal al consumo resulta de la redistribución: *c' = c1 * f1' + c2 * f2´ = c1 * [f1 * (1 – p) + p] + c2 * f2 * (1 – p) = (c1 * f1 + c2 * f2) * (1 – p) + c1 * p.* Reacomodando un poco y usando (3.3) resulta

$$c´ = c + p * (c1 - c) \qquad (3.5)$$

lo cual da un multiplicador

$$m' = 1/(1 - c') = 1/(1 - c - p * (c1 - c)) \qquad (3.6)$$

a partir del cual se encuentra la nueva producción *Y'* por la fórmula usual (2.6), concretamente

$$Y' = m' * (Iv + G + Xn) = (Iv + G + Xn) / (1 - c - p * (c1 - c)) \qquad (3.7)$$

Supongamos por un momento que el sector 1, los particulares de bajos ingresos, ganan tan poco que no tienen capacidad de ahorro. Esto implica *c1 = 1* (consumen el 100% de su ingreso). Entonces el denominador de (3.6) vale *1 – c – p * (1 – c) = (1 – c) * (1 – p)*. El multiplicador es *m' = 1 / ((1 – c) * (1 – p))*, y como *1/ (1 – c) = m* tenemos

$$m' = m / (1 - p) \qquad (3.8)$$

El multiplicador (y con él el tamaño de la torta) **aumenta exactamente en la proporción necesaria para mantener constantes los ingresos de los sectores 2 y 3**. En efecto, *Y2´= f2´ * Y´= f2 * (1 - p) * m´ * (Iv + G + Xn) = f2 * (1-p) * (m/ (1-p)) * (Iv + G + Xn)* que simplificando los *(1-p)* da *Y2´= f2 * m * (Iv + G + Xn) = f2 * Y = Y2* y similarmente para el sector 3. **Se puede** dar más a un sector sin quitarle nada a los otros, siempre y cuando el primero tenga capacidad de ahorro nula. Este sería el procedimiento institucionalmente aceptable para poner en práctica la recomendación del marciano de la sección 1.1.1. Es también la confirmación, ahora avalada por herramientas macroeconómicas serias, de la viabilidad de la "solución solidaria" hallada en el modelo de la isla.

Nos falta sin embargo resolver varios aspectos concretos. Primero, cómo identificar un sector 1 con la propiedad requerida. Segundo, cuáles son los números reales involucrados. Y tercero, cómo inducir la transición al estado final deseado. Pasamos a considerar estas cuestiones.

### 3.1.1 Definición del sector de bajos ingresos. Estimaciones numéricas

La elección más obvia es formar este sector con los desocupados, más todos los jubilados y asalariados que cumplan algún criterio razonable de "bajos ingresos". Por ejemplo, que ganen menos $1.000 mensuales si mantienen una familia, o de $500 si viven solos. Eventualmente podría restringirse más el "ingreso al club 1" poniendo condiciones adicionales (digamos, que no posean vivienda propia) sutilezas en que no entraremos. Digamos, solamente, que mayores restricciones no cambian el argumento. No importa que algunos "no ahorristas" queden fuera del sector 1, mientras los que están satisfagan la condición, lo dicho en la sección anterior vale.

Hagamos algunos números "a ojo de buen cubero" para tener idea de los órdenes de magnitud (errores de un 20 o 25% en los valores que siguen son admisibles, no cambiarían substancialmente las conclusiones). Cantidad de agentes económicos en el sector: 12.000.000 (7.000.000 de asalariados, 3.500.000 de jubilados, 1.500.000 de desocupados). Ingreso promedio $400 mensuales (aunque el asalariado típico gane más, los desocupados y jubilados tiran el promedio abajo). Ingreso del sector *Y1= 4.800* M$ mensuales, o 62.400M$ anuales (incluye aguinaldo). Fracción del ingreso [2] *f1= Y1 / Y= 62,4 / 281= 22%.*

Según la sección 2.6, el desempleo se soluciona con un incremento del 20% en el PBI, o sea 56.000M$ anuales. Bajo la hipótesis *c1= 1* y teniendo en cuenta la ecuación (2.8), se necesita *1/(1 - p) = 1,2*, lo que da *p= 0,166*. El aumento del ingreso va como vimos integramente al sector1. *Y1* pasaría de 62.000M$ anuales a 118.000M$, un aumento del 90%. El ingreso promedio de sus agentes pasa de $400 a $760, valor mucho más razonable.

Como gimnasia contra la *innumeracy*[3] y sin pretensión de exactitud, veamos que representa el anterior aumento para los no desocupados. Digamos que un trabajador en actividad cobra 2,5 veces lo que un jubilado. El *Y1* inicial de 4.800M$ se reparte entonces así: 7.000.000 de asalariados a un sueldo promedio de $572 = 4.004M$; 3.500.000 jubilados cobrando en promedio $228= 798M$; total 4802M$ (el 2 es error de redondeo). El *Y1´* final de 9.120M$ se reparte en: 8.500.000 asalariados a $921 de promedio= 7828M$; 3.500.000 jubilados a $368 = 1.288M$. Los sueldos y jubilaciones existentes suben un 61%, el otro 29% del aumento en *Y* se va en sueldos de los ex-desocupados.

---

[2] Tomamos como estado inicial el descripto en la tabla 2.1. Dada la persistencia del desempleo y la estabilidad a lo largo de los meses de otros indicadores, es lícito suponer que es aproximadamente un estado de equilibrio.

[3] El término *innumerate* fue propuesto por Martín Gardner, quien por muchos años escribió la sección Juegos Matemáticos de Scientific American, para designar a alguien incapaz de manejar y/o desarrollar intuición sobre relaciones numéricas. Se inspira en *illiterate*, iletrado o analfabeto en inglés.

Hasta acá nos hemos apoyado fuertemente en la suposición de que *c1= 1*, es decir que los integrantes del sector 1 lisa y llanamente no ahorran. ¿Qué cambia si se nos "infiltran" algunos ahorristas? En la práctica probablemente nada en el corto plazo. Estamos suponiendo un sector con necesidades de consumo largamente postergadas. De acceder a un mayor ingreso, muchos tomarán créditos para satisfacerlas. El ahorro total del sector, aún con algunos ahorristas "colados" debe ser incluso negativo (c1 > 1, consumo presente mayor que ingresos).

Hagamos sin embargo una estimación. Digamos que de cada 5 personas en el sector hay una con capacidad de ahorrar el 20% de su sueldo. Significa que de cada $500 que recibe el sector se ahorran $20. Se consumebn entonces $480, lo que da *c1= 480/500 = 0,96*. Podemos ahora asignarle valores a *p* e ir estudiando la variación de *c´, m´* e *Y´* mnediante las ecuaciones (3.5) a (3.7) Los resultados del cálculo se presentan en la tabla 3.1.

| p% | c´ | m´ | Y´ | f1´ | Y1´ | Y2´+Y3´ | Cambios porcentuales | | |
|---|---|---|---|---|---|---|---|---|---|
|  | ec. 3.5 | ec. 3.6 | ec. 3.7 | ec. 3.4.3 | f1´*Y´ | Y´- Y1´ | en Y´ | en Y1´ | Y2´+Y3´ |
| 0% | 0.690 | 3.23 | 281 | 0.22 | 62 | 219 | 0.0% | 0.0% | 0.0% |
| 2% | 0.695 | 3.28 | 286 | 0.24 | 67 | 218 | 1.8% | 9.0% | -0.3% |
| 4% | 0.701 | 3.34 | 291 | 0.25 | 73 | 218 | 3.6% | 18.3% | -0.5% |
| 6% | 0.706 | 3.40 | 296 | 0.27 | 79 | 217 | 5.5% | 28.0% | -0.8% |
| 8% | 0.712 | 3.47 | 302 | 0.28 | 85 | 216 | 7.5% | 38.0% | -1.1% |
| 10% | 0.717 | 3.53 | 307 | 0.30 | 92 | 216 | 9.5% | 48.4% | -1.4% |
| 12% | 0.722 | 3.60 | 313 | 0.31 | 98 | 215 | 11.7% | 59.2% | -1.7% |
| 14% | 0.728 | 3.67 | 320 | 0.33 | 105 | 214 | 13.9% | 70.4% | -2.1% |
| 16% | 0.733 | 3.75 | 326 | 0.34 | 112 | 214 | 16.2% | 82.1% | -2.4% |
| 18% | 0.739 | 3.83 | 333 | 0.36 | 120 | 213 | 18.6% | 94.3% | -2.8% |
| 20% | 0.744 | 3.91 | 340 | 0.38 | 128 | 212 | 21.1% | 107.0% | -3.1% |

Tabla 3.1

*El Desempleo: estudio de sus causas y posibles soluciones*     79

El incremento deseado en *Y* del 20% se produce aproximadamente para *p = 19%*. Aunque ahora ya no es gratis para los sectores 2 y 3, el costo de la recuperación para ellos sigue siendo ínfimo. Renunciando a un 3% de sus ingresos se elimina el desempleo y se aumenta en 100% el ingreso del sector "pobre". La figura 2.1 presenta los datos relevantes en forma gráfica, pudiéndose observar claramente la evolución del sector 1, y el costo insignificante que acarrea para los sectores 2 y 3.

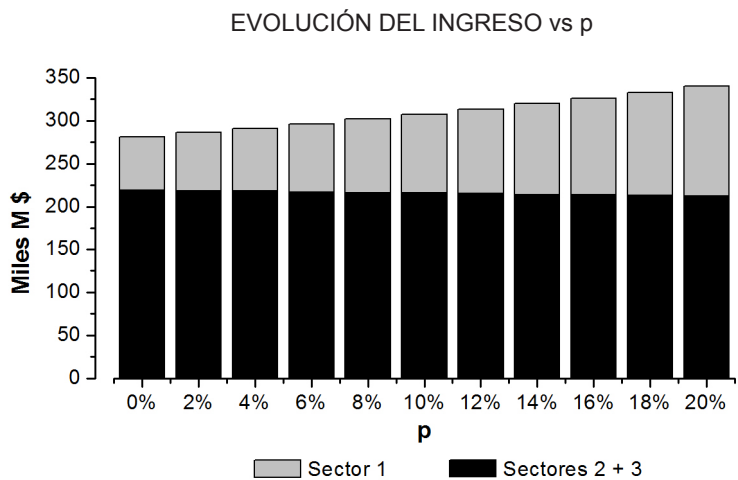

Figura 3.1

### 3.1.2 Implentación

Llevar las ideas propuestas a la práctica no es fácil porque hay que vencer una serie de hábitos mentales y prejuicios. Pero creo que es posible, y la magnitud de los potenciales beneficios justifica con creces cualquier esfuerzo que se haga. En mi opinión (abierta a críticas y sugerencias) los pasos requeridos son:

• Generar un debate en la profesión económica. Dada la naturaleza humana no puede esperarse consenso: siempre habrá quien rechace cualquier idea nueva, con o sin argumentos para oponerle. Pero se podrá: a) convencer a algunos, b) verificar que no estemos cometiendo errores ni falacias, c) encontrar cabos sueltos en base a las críticas, y corregirlos, d) ampliar las ideas con las sugerencias que se reciban.

• Una vez depurada y enriquecida la teoría hacer una amplia labor didáctica, apuntando sobre todo a dirigentes de empresas líderes y políticos. Mientras las mentes de quienes dirigen no superen la "falacia de la torta fija" o las explicaciones simplistas "el desempleo es un fenómeno mundial, es por los robots[4], no se puede hacer nada", efectivamente no se hará nada.

• "Venderle" la idea al gobierno. Es el único que puerde implementar una política de este tipo mediante incentivos y castigos fiscales. (podría hasta estudiarse la inclusión total o parcial del gobierno en el sector 1 con miras a reducir su déficit. Aunque no es "pobre", cumple el único requisito esencial para este sector: no ahorra).

• El paso del equilibrio actual a la solución solidaria tiene que ser necesariamente gradual para evitar "sacudidas". Si se requiere un *p* del 19% habrá que procurar cambios de, digamos, 1% mensual hasta llegar al nivel deseado. Adicionalmente, el avance gradual y la confrontación de resultados con previsiones permitirá efectuar sobre la marcha la sintonía fina necesaria.

• Lo importante es lograr la adhesión de las 20 ó 50 empresas mas importantes del país. Si éstas comienzan lenta y simultáneamente a subir el salario real y a aumentar el empleo, la rueda se pone en movimiento y el resto las seguirá. Una *posible* política fiscal: a) fijar pautas sobre empleo y salario para las empresas, b) reducir impuestos para las que cumplen. La reducción de las tasas será gradual, y cuantificada de tal manera que los ingresos del gobierno acompañen al resto del sector 3 (ingresos aproximadamente constantes, rebajas compensadas por el crecimiento)

• El seguimiento de 20, 50 ó 100 grandes empresas es *mucho* más fácil que la persecución impositiva de cien mil quiosqueros y mecánicos de bicicletas que efectúa actualmente la DGI. El control de las metas no presentaría mayores problemas.

En resumen, el gobierno debería juntar a los empresarios y decirles "Sres., necesitamos que este bimestre como contribución al crecimiento aumenten los salarios mas bajos en 7.2%[5] sin trasladarlo a los precios. El mayor poder adquisitivo de la gente les hará crecer las ventas en 1.8% con lo cual el resultado para Uds. es neutro. El Gobierno, para predicar con el ejemplo, aumenta igualmente sueldos y jubilaciones bajas, y a la s empresas que cumplan les dará un crédito fiscal de 1,8% sobre los impuestos que paguen. Es esto, o la agudización de la situación actual: recesión, marginalidad, conflictos sociales, huelgas..."

---

4 Otros "culpables" populares son: las multinacionales, los chinos que trabajan por un plato de arroz, el FMI, la sinarquía internacional, ...
5 Dato coherente con la fila 2% de la tabla 3.1. El 7,2% mas el 1,8% que surge del mayor empleo da el 9% de la columna *Y1´*.

Al cabio de 2 ó 3 meses evaluar los resultados, y repetir el procedimiento si fueron los esperados. De lo contrario diagnosticar los motivos (por ejemplo la mayoría de las empresas no cumplió) y corregir.

## 3.2 Mediano plazo

Independientemente de cualquier incerteza sobre el valor del multiplicador, la eduación (2.6) muestra que un aumento de la inversión *Iv* hace crecer la producción y por lo tanto el empleo. La cuestión es cómo inducir un mayor *Iv*. Existe en todo momento una serie de posibilidades de inversión en grandes obras, con buenas tasas de retorno a largo plazo, que las empresas privadas no encaran porque horizontes de tiempo tan largos no les interesan. Ejemplos para el caso argentino: forestación, regadío de tierras fértiles pero sin agua (San Luis, Patagonia, etc.), autorutas Bs. As.-Mendoza, Bs.As.- Mar del Plata (o Ushuaia?), electrificación rural. Las posibilidades son muchas, hay que explorarlas. En otros lugares y/o épocas, estas grandes obras eran encaradas por el Estado. Lo cual hoy, dadas las penurias fiscales, es impensable. ¿Qué se puede hacer al respecto?. Encuauzar el excedente de ahorro privado en esta dirección.

La propuesta es formar sociedades anónimas ad-hoc que podríamos llamar "Sociedades Siglo XXI", con un régimen jurídico propio, para este fin. Veamos como funcionarían en un caso concreto: la forestación. El Estado aporta tierras fiscales y elabora un pre-proyecto evaluando factibilidad, duración, tasas esperadas de retorno, etc. a cambio de una parte del paquete accionario. Prácticamente no le cuesta nada si lo hace con el personal que ya tiene por ejemplo en las universidades. A partir de allí la sociedadd es administrada en forma privada, con facultades de control para el Estado en su carácter de accionista. Por licitación o lo que sea, se pasa el proyecto a un grupo privado que aportará el resto del capital inicial. De allí en más se financia el funcionamiento con emisión de acciones.

Las acciones no serán del tipo convencional, abstracto, sino que ofrecerán al inversor la garantía de algo tangible. Y deberán tener "gancho". "Sr., invierta hoy $500 en Pinos y Eucaliptos Sociedad Siglo XXI, y sea dueño de cien árboles en la plantación... (indicar lugar) Nosotros se los cuidaremos durante los próximos 20 años, al cabo de los cuales Ud. o sus hijos recibirán el fruto de su comercialización (o podrán llevárselos si lo prefieren). Valor de mercado hoy: $6000, rendimiento: 13,23% anual sobre inflación. Para más detalles, solicite copia del proyecto o plantéenos sus dudas al teléfono.....", Si se hace con seriedad, una inversión de este tipo es muchísimo más ventajosa que tener dólares en una cuenta suiza o plazo fijo y debería atraer ahorristas. **Si se hace con seriedad.**

El marco jurídico para estas sociedades debe garantizar al inversor total transparencia en el manejo de **sus** ahorros, con exigencias al menos similares a las que se imponen a sociedades cotizadas en la Bolsa. Adicionalmente deberáin disponerse medidas fiscales de fomento a estas inversiones. Ejemplo: el 50% de lo invertido es sociedades siglo XXI se toma como crédito sobre el impuesto a las ganancias. Mucha gente preferirá desembolsar $5000 en 1000 árboles **propios**, a "regalarle" $2500 de impuesto al Estado y no verlos nunca más. La menor recaudación deberá lógicamente componesarse de alguna forma, siendo la más razonable "mano dura" con los actuales evasores. De todos modos, una pequeña caída de la recaudación es tolerable: si cae el desempleo, el gobierno reduce su gasto social.

Finalmente, la iniciativa propuesta tiene otra gran virtud. No sólo reduce el desempleo, también hará que le dejemos un país más próspero a las generaciones futuras.

## 3.3 Perspectivas a largo plazo

Creo que la única organización social posible a largo plazo es el reemplazo de la competencia por la solidaridad. A medida que la tecnología avanza la productividad crece. En la figura 3.2 hemos reproducido la 1.3 con leves modificaciones, en particular con *Ventas* reemplazada por el concepto correcto de *Demanda Efectiva*. Las líneas "Producción 1, 2 y 3¨muestran la misma como podría ser en los años 1990, 2010 y 2030. Como cada vez se produce más con la misma cantidad de mano de obra, las gráficas de producción se vuelven más y más empinadas tendiendo a confundirse con el eje vertical.

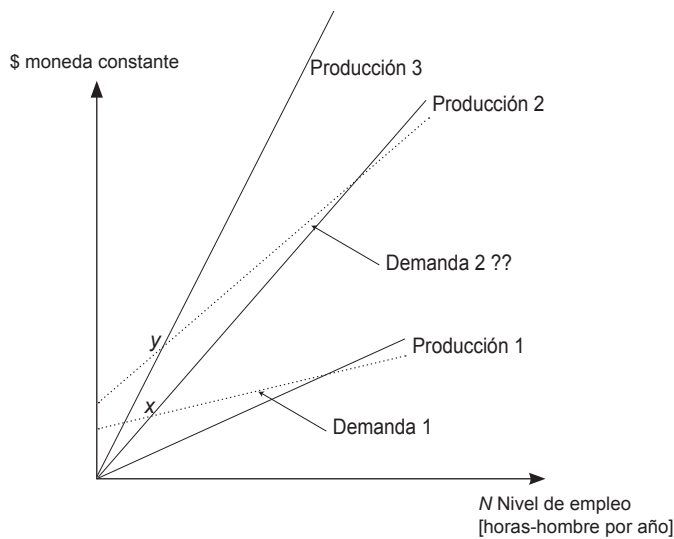

Figura 3.2

Si la demanda no cambia mucho entre 1990 y 2010, el nivel de empleo cae al punto x, inaceptablemente bajo. Lo que hubiéramos necesitado es que *N* aumente acompañando el crecimiento demográfico. En lugar de eso cayó drásticamente. Supongamos sin embargo que por algún suceso excepcional que genera inversión (colonización de la luna?) se consigue llegar al 2010 con la demanda 2, manteniendo mas o menos el *status quo* en cuando a empleo. De nada sirve, en otros 20 años el equilibrio habrá caído a *y*. Estamos como el corredor que pretende alcanzar a su sombra.

Conclusión: con la organización social actual las alternativas son

• Mantener el pleno empleo mediante una orgía de producción que desafía al sentido común, y seguramente a la capacidad ecológica del planeta. Producción que, por otra, depende de circunstancias excepcionales para ser absorbida.

• Mantener un nivel de producción razonable con un desempleo que tiende inexorablemente al 100%!!

En realidad, para detectar que **algo** debe cambiar, ni nos hace falta el gráfico. La máquina sustituye cada vez más al hombre en el proceso de producción. ¿Qué pasará el día en que los robots hagan **todo** el trabajo? Sus dueños tendrán satisfechas sus necesidades materiales. ¿Y el resto? ¿Vivirá de la caza y de la pesca? Imposible, los bosques y costas tendrán dueño, no hay dónde cazar ni pescar.

Sin ser tan drásticos ¿qué pasará el día que los robots hagan el 95% del trabajo humano actual? Un operario en 8 hs. producirá lo mismo que 20 operarios hoy. ¿Se mantendrá la producción actual echando a los 19 restantes? ¿O se producirá 20 veces más? ¿Para quién? Este escenario no es de ciencia ficción, al paso vertiginoso de la técnica, un factor 20 en la productividad puede darse en pocas décadas.

### 3.3.1 ¿Cuál es la solución?

La única posibilidad a largo plazo que veo es la *solución solidaria*: **repartir** los frutos de la tecnificación. Si en el punto *x* se requiere un 12-avo de la fuerza laboral disponible, que cada uno trabaje un mes por año pero *cobrando los 12.* El resto del tiempo podrá dedicarlo al ocio o a actividades creativas, estudio, artes, deportes, ciencias, etc. Florecerán los Mozarts y los Maradonas, los filatelistas y jugadores de truco, habrá 100 veces más médicos que hoy investigando el cáncer. Y si en otros 20 años se está en el punto *y*, pues que cada uno trabaje 15 días.

La solución solidaria significa el abandono del modelo competitivo, no hay parche que pueda hacerlos compatibles. Estamos diciendo que cuanto menos sea la demanda de mano de obra, mejor hay que pagarla. Es la *negación* de la filosofía del mercado.

El paso a la solución solidaria por supuesto requiere cambiar totalmente las actitudes y valores sociales. Mientras se valore a las personas por lo que poseen no hay chance. El Eduardo de nuestra isla puede muy bien estar en contra de la solución solidaria aunque reciba los mismos pescados porque pierde su *privilegio relativo.* Una cosa es ser dueño de un auto, otra cosa es tener el único auto del pueblo. Pero a largo plazo creo que es: o solidaridad o caos.

Hace 2000 años, en medio de la inequidad y el egoísmo, un hombre predicó el amor al prójimo y revolucionó el mundo. La humanidad deberá decidir, y pronto, si sigue o no sus enseñanzas.